# Performance of the front-end electronics of the ANTARES neutrino telescope

ANTARES Collaboration

J.A. Aguilar [a], I. Al Samarai [b], A. Albert [c], M. Anghinolfi [d], G. Anton [e], S. Anvar [f], M. Ardid [g], A.C. Assis Jesus [h], T. Astraatmadja [h,1], J-J. Aubert [b], R. Auer [e], B. Baret [i], S. Basa [j], M. Bazzotti [k,l], V. Bertin [b], S. Biagi [k,l], C. Bigongiari [a], M. Bou-Cabo [g], M.C. Bouwhuis [h], A. Brown [b], J. Brunner [b,2], J. Busto [b], F. Camarena [g], A. Capone [m,n], L. Caponetto[ae,3], C.Cârloganu [o], G. Carminati [k,l], J. Carr [b], E. Castorina [p,q], V. Cavasinni [p,q], S. Cecchini [l,r], Th. Chaleil [f], Ph. Charvis [s], T. Chiarusi [l], N. Chon Sen [c], M. Circella [t], H. Costantini [d], N. Cottini [u], P. Coyle [b], C. Curtil [b], G. De Bonis [m,n], N. de Botton [u], I. Dekeyser [v], E. Delagnes [f], A. Deschamps [s], C. Distefano [w], C. Donzaud [i,x], D. Dornic [b,a], D. Drouhin [c], F. Druillole [f], T. Eberl [e], U. Emanuele [a], J-P. Ernenwein [c,b], S. Escoffier [b], E. Falchini [p,q], F. Fehr [e], F. Feinstein [u,b,4], V. Flaminio [p,q], J. Fopma [5], K. Fratini [y,d], U. Fritsch [e], J-L. Fuda [v], P. Gay [o], G. Giacomelli [k,l], J.P. Gómez-González [a], K. Graf [e], G. Guillard [z], G. Halladjian [b], G. Hallewell [b], C. Hoffmann [z], H. van Haren [aa], A.J. Heijboer [h], Y. Hello [s], J.J. Hernández-Rey [a], B. Herold [e], J. Hößl [e], M. de Jong [h,1], N. Kalantar-Nayestanaki [ab], O. Kalekin [e], A. Kappes [e], U. Katz [e], P. Kooijman [h,ac,ad], C. Kopper [e], A. Kouchner [i], W. Kretschmer [e], D. Lachartre [f,6], H. Lafoux [u], R. Lahmann [e], P. Lamare [f], G. Lambard [b], G. Larosa [g], H. Laschinsky [e], H. Le Provost [f], A. Le Van Suu [b,7], D. Lefèvre [v], T. Legou [b,4], G. Lelaizant [b], G. Lim [h,ad], D. Lo Presti [ae], H. Loehner [ab], S. Loucatos [u,*], F. Lucarelli [m,n], S. Mangano [a], M. Marcelin [j], A. Margiotta [k,l], J.A. Martinez-Mora [g], A. Mazure [j], E. Monmarthe [f], T. Montaruli [t,af], M. Morganti [p,q], L. Moscoso [u,i], H. Motz [e], C. Naumann [u], M. Neff [e], Ch. Olivetto [b,i], R. Ostasch [e], D. Palioselitis [h], G.E.Păvălaș [ag], P. Payre [b], J. Petrovic [h], P. Piattelli [w], N. Picot-Clemente [b], C. Picq [u], J.-P. Pineau [z], J. Poinsignon[f], V. Popa [ag], T. Pradier [z], E. Presani [h], C. Racca [c], A. Radu [ag], C. Reed [b,h], F. Réthoré [b,j], G. Riccobene [w], C. Richardt [e], M. Rujoiu [ag], G.V. Russo [ae], F. Salesa [a], P. Sapienza [w], F. Schöck [e], J P. Schuller [u], R. Shanidze [e], F. Simeone [n], M. Spurio [k,l], J.J.M. Steijger [h], Th. Stolarczyk [u], C. Tamburini [v], L. Tasca [j], S. Toscano [a], B. Vallage [u], V. Van Elewyck [i], G. Vannoni [u], M. Vecchi [m], P. Vernin [u], G. Wijnker [h], E. de Wolf [h,ad], H. Yepes [a], D. Zaborov [ah], J.D. Zornoza [a], J. Zúñiga [a]

[a]*IFIC - Instituto de Física Corpuscular, Edificios Investigación de Paterna, CSIC - Universitat de València, Apdo. de Correos 22085, 46071 Valencia, Spain*

[b]*CPPM - Centre de Physique des Particules de Marseille, CNRS/IN2P3 et Université de la Méditerranée, 163 Avenue de Luminy, Case 902, 13288 Marseille Cedex 9, France*

[c]*GRPHE - Institut universitaire de technologie de Colmar, 34 rue du Grillenbreit BP 50568 – 68008 Colmar, France*

[d]*INFN - Sezione di Genova, Via Dodecaneso 33, 16146 Genova, Italy*

[e]*Friedrich-Alexander-Universität Erlangen-Nürnberg, Erlangen Centre for Astroparticle Physics, Erwin-Rommel-Str. 1, D-91058 Erlangen, Germany*

[f]*Direction des Sciences de la Matière - Institut de recherche sur les lois fondamentales de l'Univers - Service d'Electronique des Détecteurs et d'Informatique, CEA Saclay, 91191 Gif-sur-Yvette Cedex, France*

[g]*Institut d'Investigació per a la Gestió Integrada de Zones Costaneres (IGIC) - Universitat Politècnica de València. C/ Paranimf, 1. E-46730 Gandia, Spain.*

[h]*FOM Instituut voor Subatomaire Fysica Nikhef, Science Park 105, 1098 XG Amsterdam, The Netherlands*

[i]*APC - Laboratoire AstroParticule et Cosmologie, UMR 7164 (CNRS, Université Paris 7 Diderot, CEA, Observatoire de Paris) 10, rue Alice Domon et Léonie Duquet 75205 Paris Cedex 13, France*

[j]*LAM - Laboratoire d ' Astrophysique de Marseille, Pôle de l'Etoile Site de Ch-Gombert, rue Frédéric Joliot-Curie 38, 13388 Marseille cedex 13, France*

[k]*Dipartimento di Fisica dell'Università, Viale Berti Pichat 6/2, 40127 Bologna, Italy*

[l]*INFN - Sezione di Bologna, Viale Berti Pichat 6/2, 40127 Bologna, Italy*

[m]*Dipartimento di Fisica dell'Università La Sapienza, P.le Aldo Moro 2, 00185 Roma, Italy*

-------------------

* Corresponding author, s.loucatos@cea.fr

[n]*INFN -Sezione di Roma, P.le Aldo Moro 2, 00185 Roma, Italy*
[o]*Clermont Université, Université Blaise Pascal, CNRS/IN2P3, Laboratoire de Physique Corpusculaire, BP 10448, F-63000 Clermont-Ferrand, France*
[p]*Dipartimento di Fisica dell'Università, Largo B. Pontecorvo 3, 56127 Pisa, Italy*
[q]*INFN - Sezione di Pisa, Largo B. Pontecorvo 3, 56127 Pisa, Italy*
[r]*INAF-IASF, via P. Gobetti 101, 40129 Bologna, Italy*
[s]*Géoazur - Université de Nice Sophia-Antipolis, CNRS/INSU, IRD, Observatoire de la Côte d'Azur and Université Pierre et Marie Curie F-06235, BP 48, Villefranche-sur-mer, France*
[t]*INFN - Sezione di Bari, Via E. Orabona 4, 70126 Bari, Italy*
[u]*Direction des Sciences de la Matière - Institut de recherche sur les lois fondamentales de l'Univers - Service de Physique des Particules, CEA Saclay, 91191 Gif-sur-Yvette Cedex, France*
[v]*COM - Centre dOcéanologie de Marseille, CNRS/INSU et Université de la Méditerranée, 163 Avenue de Luminy, Case 901, 13288 Marseille Cedex 9, France*
[w]*INFN - Laboratori Nazionali del Sud (LNS), Via S. Sofia 62, 95123 Catania, Italy*
[x]*Université Paris-Sud 11 - Département de Physique - F - 91403 Orsay Cedex, France*
[y]*Dipartimento di Fisica dell'Università, Via Dodecaneso 33, 16146 Genova, Italy*
[z]*IPHC-Institut Pluridisciplinaire Hubert Curien - Université de Strasbourg et CNRS/IN2P3 23 rue du Loess -BP 28- F67037 Strasbourg Cedex 2*
[aa]*Royal Netherlands Institute for Sea Research (NIOZ), Landsdiep 4,1797 SZ 't Horntje (Texel), The Netherlands*
[ab]*Kernfysisch Versneller Instituut (KVI), University of Groningen, Zernikelaan 25, 9747 AA Groningen, The Netherlands*
[ac]*Universiteit Utrecht, Faculteit Betawetenschappen, Princetonplein 5, 3584 CC Utrecht, The Netherlands*
[ad]*Universiteit van Amsterdam, Instituut voor Hoge-Energie Fysika, Science Park 105, 1098 XG Amsterdam, The Netherlands*
[ae]*Dipartimento di Fisica ed Astronomia dell'Università, Viale Andrea Doria 6, 95125 Catania, Italy*
[af]*University of Wisconsin - Madison, 53715, WI, USA*
[ag]*Institute for Space Sciences, R-77125 Bucharest, Măgurele, Romania*
[ah]*ITEP - Institute for Theoretical and Experimental Physics, B. Cheremushkinskaya 25, 117218 Moscow, Russia*

[1] *Also at University of Leiden, the Netherlands.*
[2] *On leave at DESY, Platanenallee 6, D-15738 Zeuthen, Germany.*
[3] *Now at IPNL Lyon, France.*
[4] *Now at L.P.T.A., Université Montpellier II/CNRS-IN2P3, CC 70, Bât. 13, place Eugène Bataillon, 3409 Montpellier Cedex 5, France.*
[5] *Central Electronics Group, Department of Physics, University of Oxford, Keble Road, Oxford OX1 3RH, UK.*
[6] *Now at CEA / Léti, 17 rue des Martyrs, 38054 Grenoble cedex 9, France.*
[7] *Now at LPL - Laboratoire Parole et Langage, UMR 6057 (CNRS, Université de Provence) 5, avenue Pasteur, BP 80975 13604 Aix-en-Provence Cedex 1, France.*

## Abstract

ANTARES is a high-energy neutrino telescope installed in the Mediterranean Sea at a depth of 2475 m. It consists of a three-dimensional array of optical modules, each containing a large photomultiplier tube. A total of 2700 front-end ASICs named Analogue Ring Samplers (ARS) process the phototube signals, measure their arrival time, amplitude and shape as well as perform monitoring and calibration tasks. The ARS chip processes the analogue signals from the optical modules and converts information into digital data. All the information is transmitted to shore through further multiplexing electronics and an optical link. This paper describes the performance of the ARS chip; results from the functionality and characterization tests in the laboratory are summarized and the long-term performance in the apparatus is illustrated.

*Keywords*: neutrino telescope; photomultiplier tube; front-end electronics; ASIC.

# 1 The ANTARES experiment

In May 2008 the ANTARES collaboration completed the deployment of a deep-sea neutrino telescope at a site located 40 km south of Toulon (France), 42°48´ N, 6°10´ E, at a depth of 2475 m [1]. The purpose of the experiment is to explore the universe using high-energy neutrinos as messengers. One of the major objectives is the identification of the accelerators producing high-energy cosmic rays. Neutrinos allow observations at distances and in a range of the energy spectrum inaccessible to photons. The potential sources of cosmic neutrinos are particle accelerators in the universe, both galactic (supernova remnants, binary stars, micro-quasars, etc.) and extragalactic (active galactic nuclei, gamma ray bursters, etc.). The indirect search for dark matter, possibly producing neutrinos by annihilation, is also one of the ANTARES goals. Beyond searches for known sources there is the possibility for completely unexpected discoveries, as has often occurred in the past with new astronomical instruments.

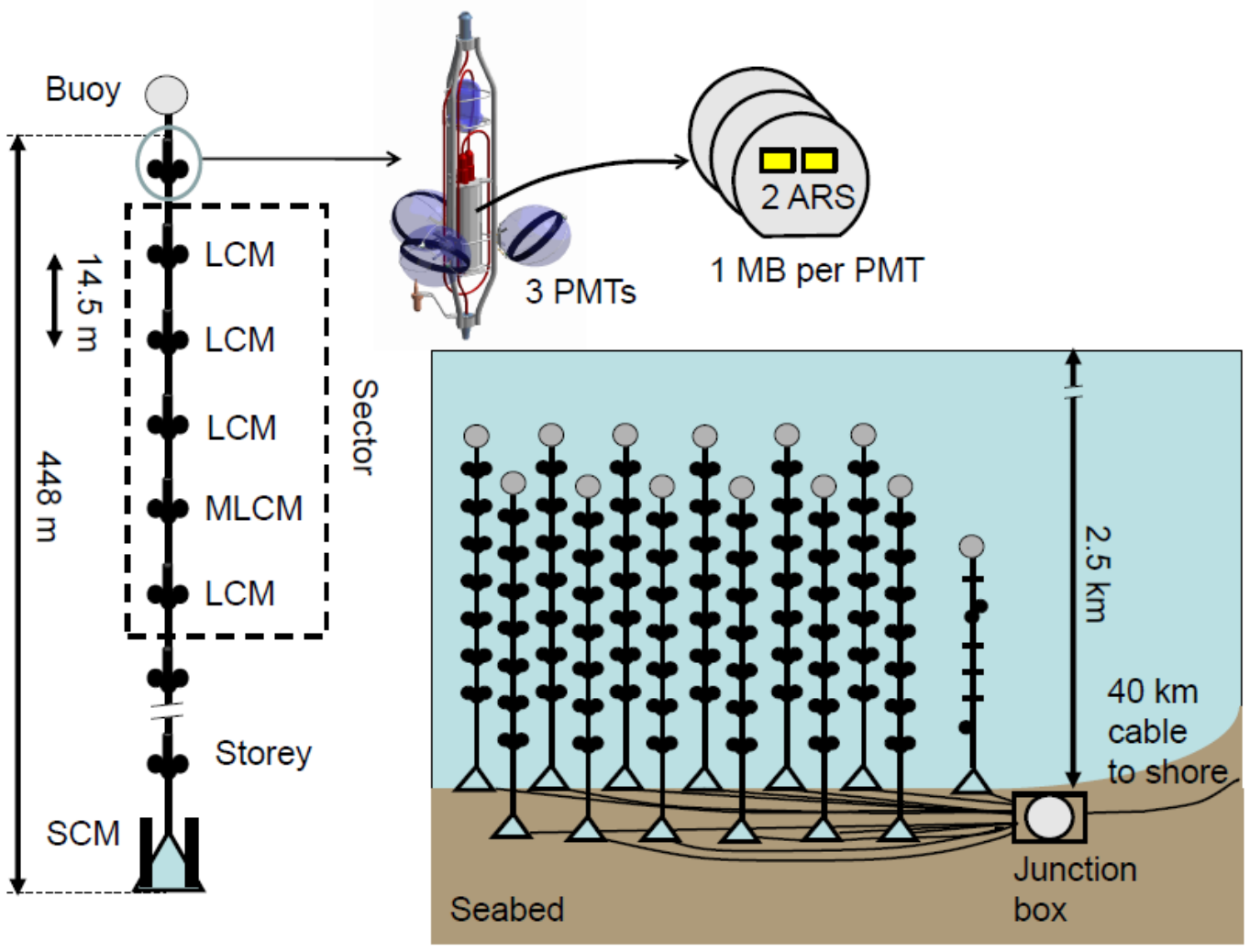

Figure 1. Schematic view of the ANTARES detector.

High-energy neutrinos may interact while crossing the Earth. Charged particles produced by the ascending neutrinos in the water or rock close enough to the apparatus will induce Cherenkov light while travelling in the water. This light is detected by a three-dimensional array of optical sensors allowing the reconstruction of long muon tracks. In this context downward atmospheric muons, the flux of which is attenuated by the column of water above the apparatus, are a source of background but may also serve for calibration purposes.

ANTARES is conceived as an array of 885 optical modules (OMs) [2] housing large-area photomultiplier tubes (PMTs) [3]. The main element of the apparatus is the line, a flexible structure built with up to 25 detection storeys connected by segments of a cable providing the mechanical strength to ensure the line integrity and distributing power and data communications through optical fibres. The lines are anchored to the sea bottom and kept taut by the buoyancy of the instrument containers and by a buoy installed at their top. Each line is connected by an underwater cable (installed by means of a submarine vehicle) to an underwater junction box, which is connected to shore by a long-distance electro-optical cable.

The full apparatus comprises 12 detection lines and one shorter instrumentation line for calibration and environmental monitoring (Figure 1). The detection lines are installed at distances of the order of 60 m from one another inside an area of about 200 m diameter. The spacing between consecutive storeys on a detection line is 14.5 m, and the bottom storey is at about 100 m above the seabed. Hence, the typical line height is approximately 450 m. One of the lines is different, since its top storeys are equipped with a prototype system of acoustic detectors [4] instead of optical modules.

A detection storey consists in a triplet of OMs. OMs are 10” hemispherical photomultipliers enclosed in pressure-resistant glass spheres. They are arranged symmetrically on a titanium frame with the photomultipliers pointing 45° downwards. The frame hosts also an electronics module and, for specific storeys of the lines, calibration or monitoring devices. The electronics module of a storey, called Local Control Module (LCM), contains all the electronics necessary for the data acquisition and control of all the devices of the storey and for communication with the shore. Each detection line is divided into 5 functional units, called sectors, each one comprising 5 storeys. The apparatus is designed so as to allow independent communications with each sector from shore. Inside each sector one electronics module, called Master Local Control Module (MLCM), acts as the gateway for all communications between its sector and the shore. An additional electronics container, the String Control Module (SCM), is located at the bottom of each line and provides the interface between each line and the rest of the apparatus.

From the point of view of the data acquisition and slow control the apparatus is conceived as an Ethernet network of 315 offshore nodes, one in each electronics module [5]. Data communications exploit a Dense Wavelength Division Multiplexing (DWDM) technique; the SCM and each MLCM of a line are equipped for optical communications using separate wavelength channels with 400 GHz spacing. The signals coming from the various sectors of the line and from the SCM are then multiplexed into a single fibre inside the SCM container. By the same token, the signals coming from the shore and directed to the different sectors of the line are de-multiplexed inside the SCM and delivered, according to their DWDM channel, to the proper destination.

The MLCMs are equipped with an Ethernet switch and with electro-optical transceivers which allow them to communicate with the LCMs of their sector in a star-configured network. This architecture allows implementing a high-rate data transfer system with a limited number of optical fibres. In this way, it is possible to transfer to shore essentially all the information collected by the apparatus, with no or minimum offshore filtering.

In order to establish a common time reference for the complete apparatus a clock signal is generated onshore, synchronized with a GPS receiver for absolute time reference and delivered to all offshore electronics by means of a dedicated optical fibre network. This network is mainly composed of passive devices for maximum reliability.

# 2 Signals from the optical modules

A natural optical background is present in deep sea water due to the Cherenkov radiation emitted by electrons originating from radioactive decays of $^{40}$K nuclei and to the light emitted by local bioluminescent organisms. The latter shows seasonal variations and sometimes large variations, even on short time scales. Such background is important as it determines the bandwidth requirement for data transmission to shore.
The default readout mode of ANTARES is the transmission of the time and amplitude of all photomultiplier signals above a threshold of 15 mV which is 1/3 of the signal of a photo-e1ectron (p.e.), 45 mV. Time measurements are referenced to the master reference clock signal distributed to each storey from shore. All signals are sent to shore and treated in a computer farm to find hit patterns corresponding to muon tracks or other physics events producing light in the water. The grouping of three optical modules in a storey allows local coincidences useful for this pattern finding and also, if required, local triggers to reduce the readout rate.

The large number of PMTs in the apparatus requires power consumption minimization for all electronics offshore. In addition, the electronics should offer large configurability options to the user and should be able to perform efficient data compression in order to minimize the offshore filtering of the data. For these reasons, the ANTARES Collaboration decided to build an ASIC, the Analogue Ring Sampler (ARS), for the front-end treatment of the OM signals. One single OM is interfaced by three such chips, installed on a motherboard. Inside each LCM a Data Acquisition (DAQ) / Slow Control board, equipped with an FPGA and a microprocessor, controls the operation of the three ARS motherboards of its storey, collects and sends their signals to shore, by means of the communications network.

# 3 The front-end motherboard

The output of each PMT goes to a motherboard. Two ARS circuits operating in a token ring configuration are connected to the output of each OM. Because of the detector size and the length of the optical link to the shore, electrical signals cannot be transmitted in analogue form; the ARS processes the analogue PMT signals and digitizes them.

The board, shown in Figure 2, is the interface between the PMT and the DAQ system. It supports two ARSs for signal processing and one ARS circuit used for additional functions. A standard LCM contains 3 motherboards for the 3 OMs of its storey. In some LCMs a 4$^{th}$ board is also present for the processing of the signals of the 8 mm PMT monitoring the LED beacons used for time calibration (Section 6.3).

A transformer is used for impedance matching of the PMT signal as shown in Figure 3. The PMT can be considered as a current source; hence the best way to transmit the signal is in differential current mode between the anode and the last dynode. The current is converted into a voltage when going to the motherboard by means of an input impedance of 25 Ω. The cable impedance is 100 Ω and so the transformer is used to match impedances, avoiding reflection and distortion of the signal. A single p.e. from the PMT, operated at a gain of $5 \times 10^7$ yields a 45 mV pulse signal on the 25 Ω output impedance after the transformer.

The digitized data of the ARS go to the DAQ board for storage, multiplexing and transmission to shore. All intermediate communication signals during the acquisition phase are in DCL[1], which is a 50 µA current logical level implemented in the ARS to avoid perturbations during data taking. This low current can only generate small voltage perturbations and thus cross-talk between internal and external signals is minimized. The motherboard ensures the conversion between DCL and the LVDS[2] signals needed for communications with the DAQ board.

[1] Direct Current Logic

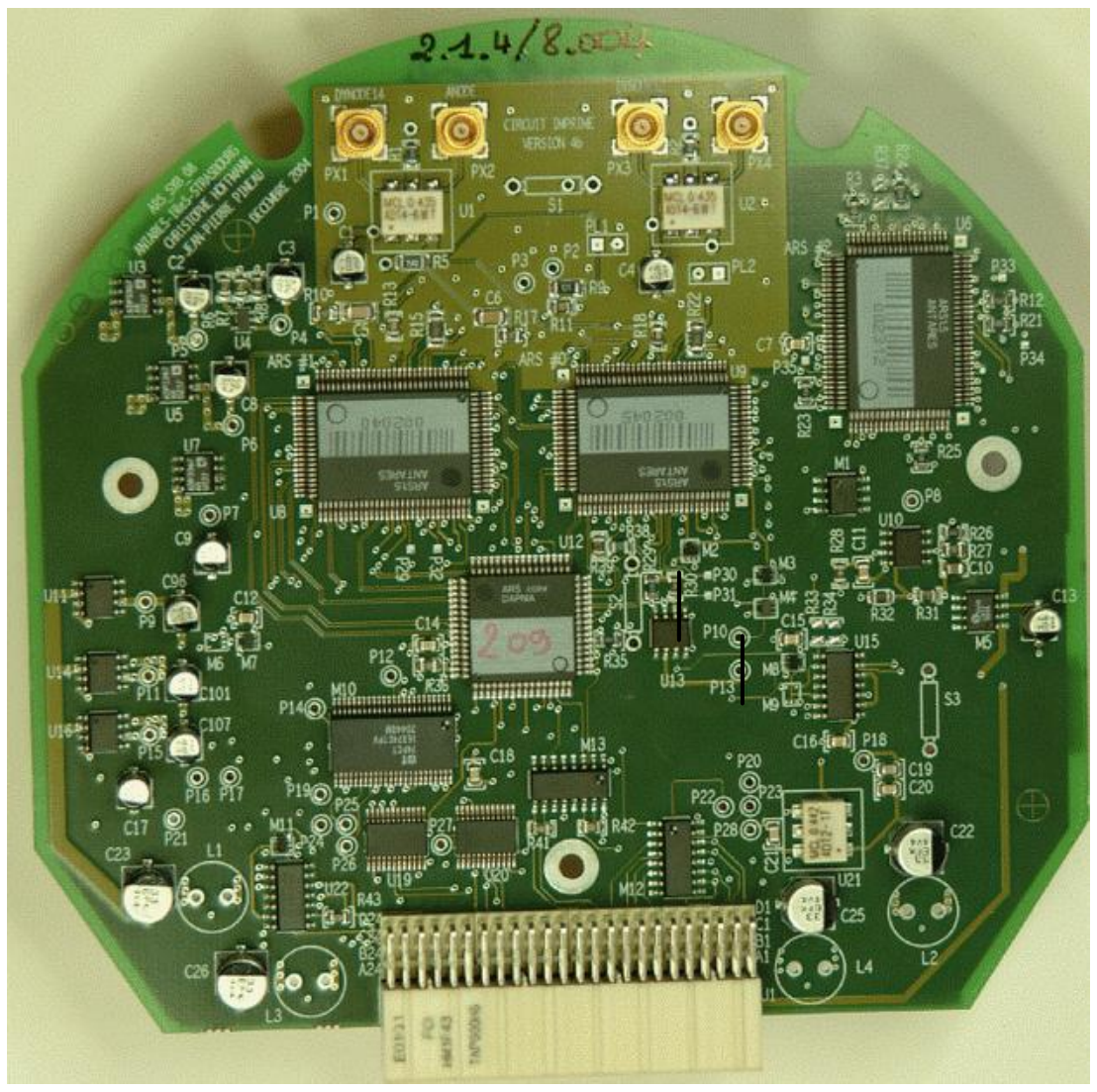

Figure 2. The ARS motherboard (diameter ~15 cm).

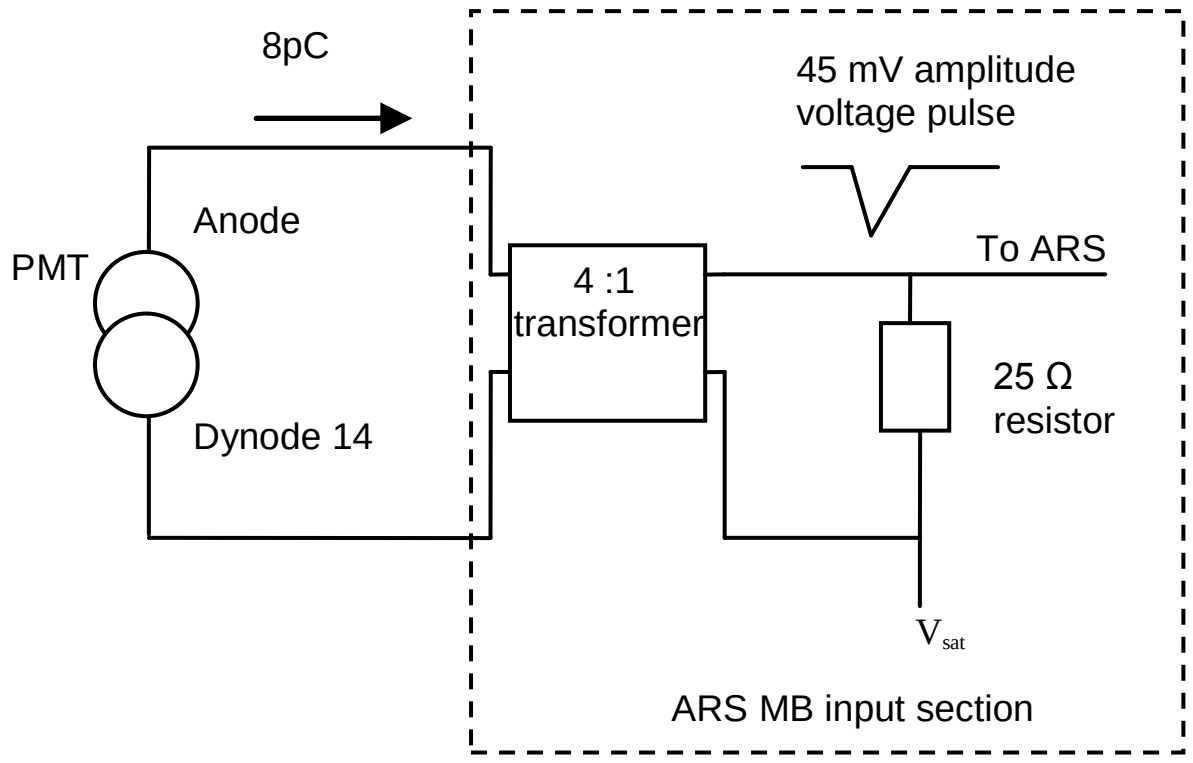

Figure 3. Impedance matching between the PMT and the ARS chips on the ARS motherboard.

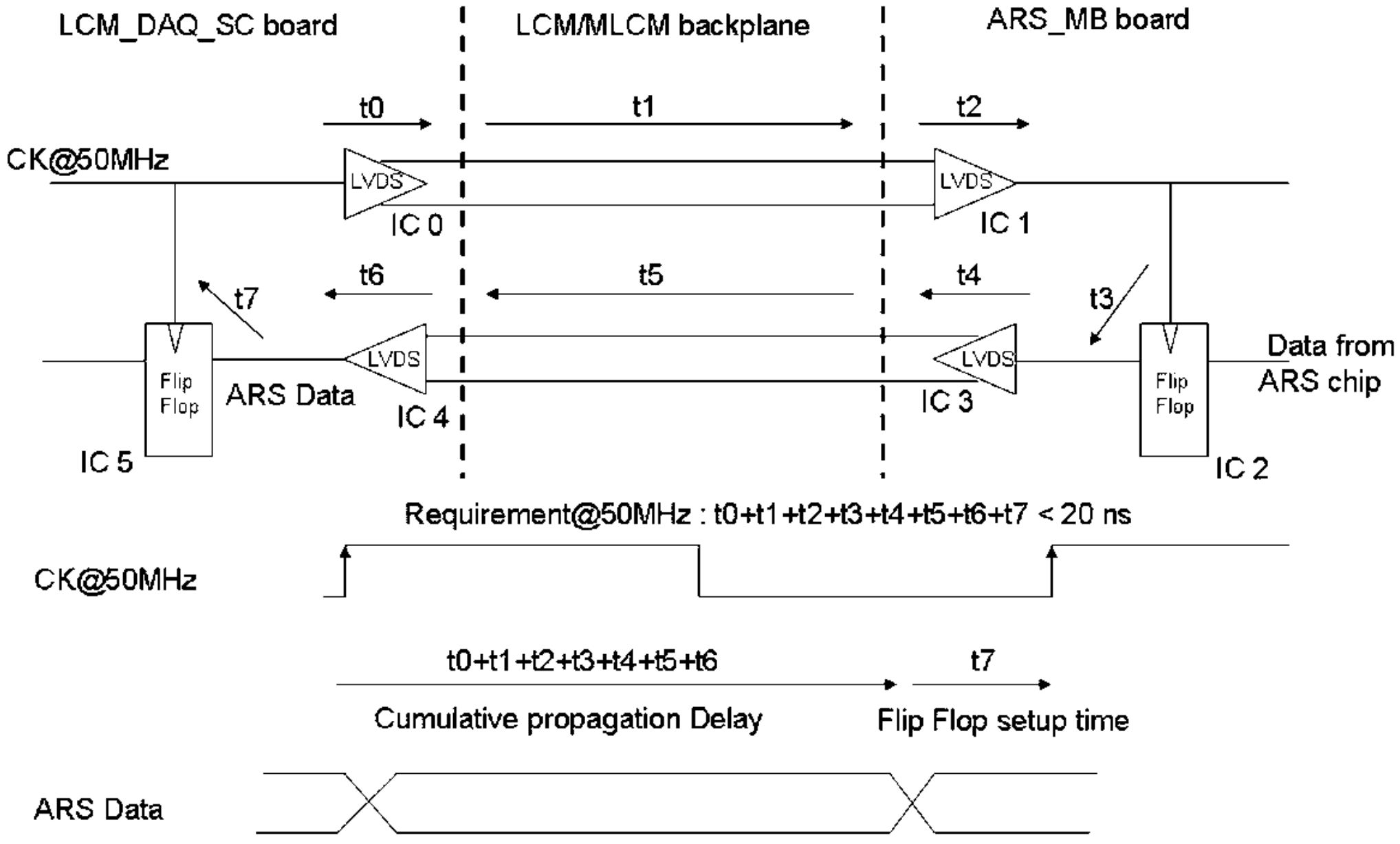

Figure 4. Data communication between the DAQ (left) and the ARS motherboard (right).

The readout is driven by a local clock of 50 MHz generated on the DAQ board. A lower frequency of 25 MHz is sent to the ARS, which uses both the rising and falling edges of the pulses to send back the associated data. Due to the protocol overhead the effective data transfer rate is 25 Mbits/s. A delicate operation performed by the motherboard is the resynchronization of the data sent to the DAQ board. It implies severe constraints on the electronic path between the DAQ board and the ARS motherboards, as shown in Figure 4. The cumulative delay in the worst case configuration reaches 19.1 ns for a 20 ns period of the clock signal. This led to the choice of fast components like the IDT74FCT16374ET[3] and FIN1531M[4] with propagation delays lower than 2 ns.

---

[2] Low Voltage Differential Signaling

[3] LVDS 4-bit high speed differential driver by Fairchild Semiconductor

The motherboard provides other functions such as the generation of a pulsed signal which can be sent to the OM to trigger an internal LED for calibration purposes, or a high-threshold trigger. These functions are based on the usage of the third ARS circuit on the motherboard, since its Digital-to-Analogue Converters (DACs) can be easily programmed to provide static voltage control.

The motherboard ensures the transmission of the slow control via the DAQ system to the ARS chips for their configuration. Because of the complexity of the design and the number of integrated components the parameters of the main electronic blocks are set via a 239-bit scan path serial link. In order to program the registers the complete bit frame of 239 bits is sent serially to all the registers through a dedicated path and all the bits are rewritten each time.

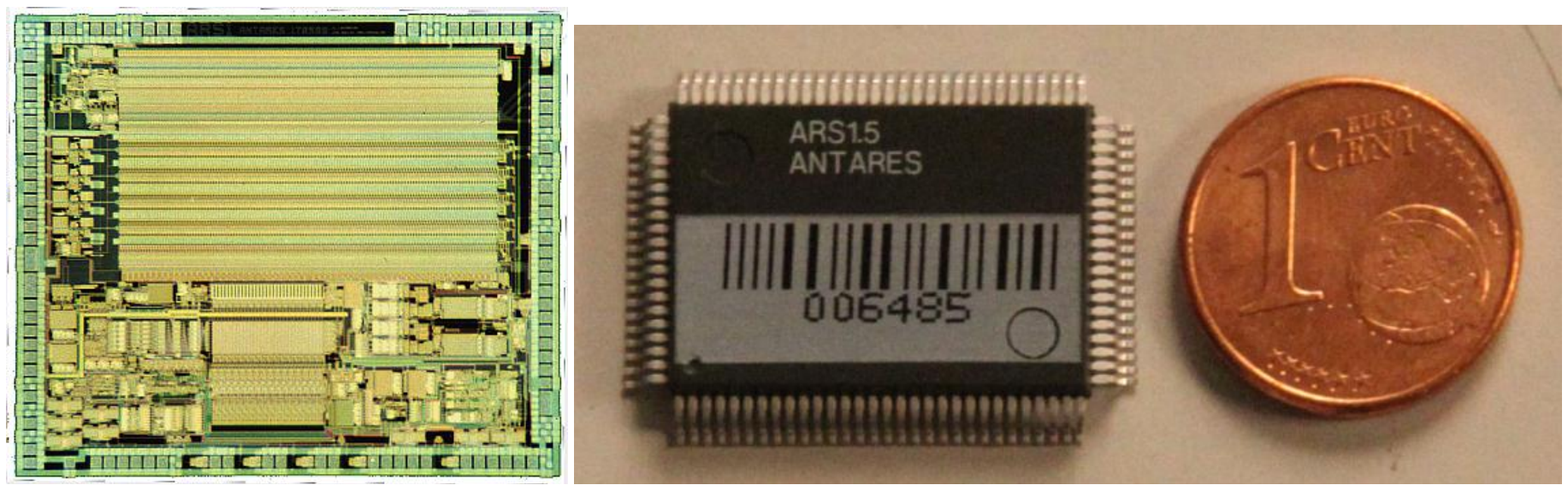

Figure 5. Images of the ARS chip, bare die (left) and packaged chip (right).

# 4 The ARS ASIC

The chip is also known in its final version as ARS1[5] and is based on two previous circuits ARS0 and ARS_SPE [6]. It is a 0.8 µm CMOS AMS technology[6] circuit, containing 68,000 transistors. It is fed by a monorail power supply 0-5V. It is described in more detail in [7,8] and a user manual provides the operation instructions [9]. Figure 5 shows pictures of the chip, whose architecture is illustrated in Figure 6.

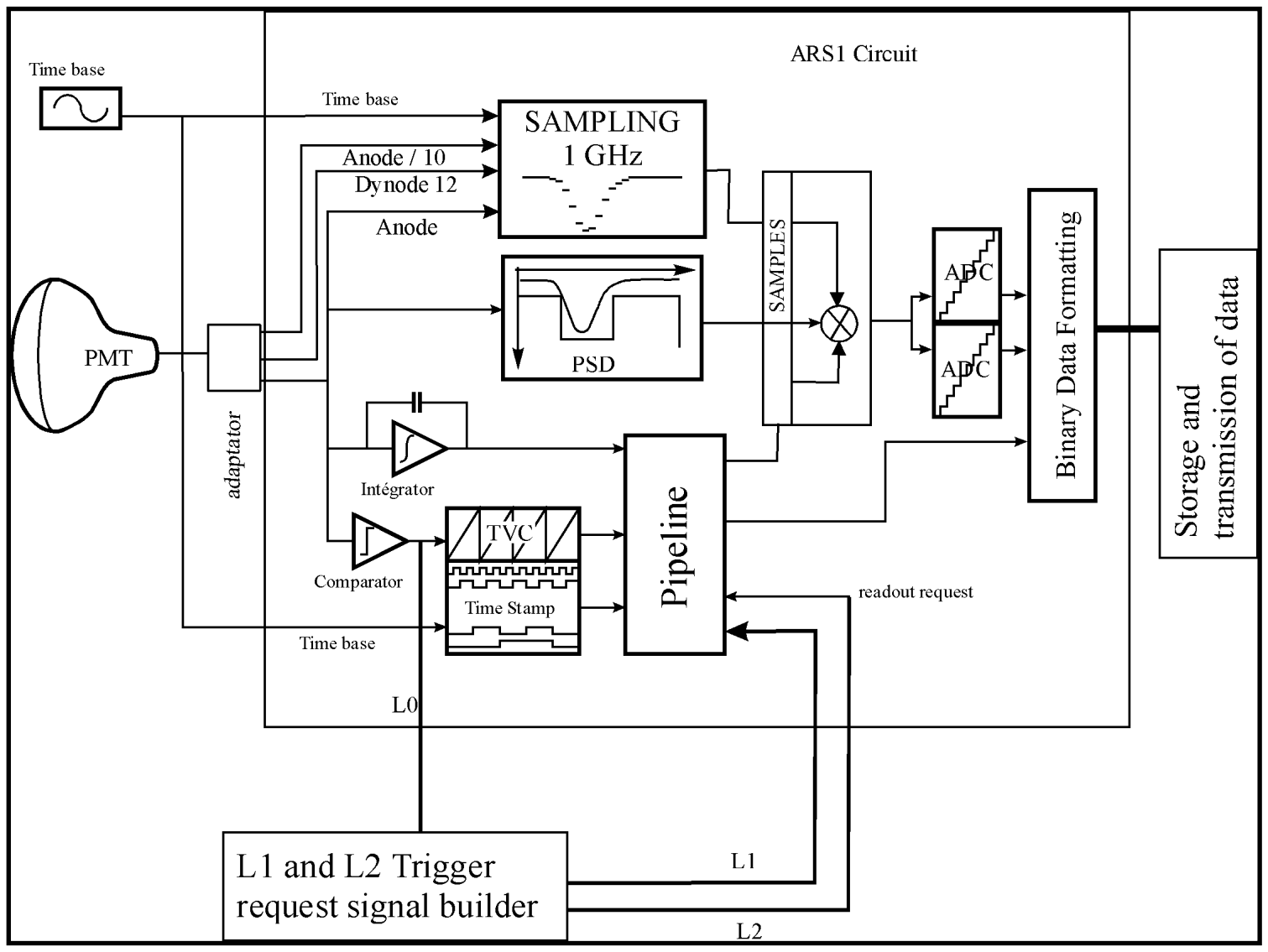

Figure 6. The ARS architecture.

[4] 16-bit fast CMOS register by IDT
[5] The last submission version was ARS1.5
[6] Complementary Metal Oxide Semiconductor technology initiated by Austriamicrosystems

The main functionalities of the ARS are as follows:

- it can discriminate between two modes, single-photoelectron (SPE) signals (i.e., signals which have a shape and an amplitude compatible with those of a single photoelectron) and complex ("waveform"-WF) signals;
- it can measure the charge and the arrival time of the event and, for WF type events, it can sample the signal at high speed and digitize it.

The fast sampling is based on track and hold cells [10,11]. The sampler is equipped with four input channels which are sampled synchronously. The signals of the anode and dynode-12 of the PMT, the 10" Hamamatsu R7081-20, which has 14 dynodes, are connected to two channels of the sampler. Dynode 12 yields a signal about 15 times lower than the anode, allowing the analysis of large signals. For similar reasons a signal which is obtained by reducing the anode signal by a factor five is connected to the third channel of the chip. The last channel samples the 20 MHz reference clock needed for time base synchronization.

The PMT anode is also connected to a threshold comparator, to a charge integrator and to a pulse shape discriminator (PSD). The PSD contains two additional threshold comparators.

Three modes of operation can be set depending on the allowed data rate and the physics goals:

- the ARS chip records only in SPE mode. The transmissible event rate is the highest;
- it records only in WF mode. This mode allows full reconstruction of the pulse shape but at a lower event rate;
- in the discriminating mode (PSD) the circuit decides which switches between SPE and WF mode depending on the shape of the PMT pulses.

SPE is the typical acquisition mode used in ANTARES, while the WF mode is used for calibration purposes. The SPE signal processing minimizes power consumption and the transmitted data volume.

Some independent functions were added to the ARS to monitor the experiment:

- a pulse train can be sent to an external LED pulser to inject light into the OM for time calibration purposes;
- a Counting Rate Monitor (CRM) records the time needed to reach a preset number of counts and the rate is deduced. A flag can be generated when the event rate exceeds a given threshold.

## 4.1 Functional description

The ARS is an asynchronous (event-driven) circuit, driven by the pulses coming from a photomultiplier. The circuit contains 24 DACs with inputs from 3 to 8 bits for parameter setting, two 8-bit ADCs, an integrator followed by an analogue-to-voltage converter (AVC), a Time-to-Voltage Converter (TVC), a 16-memory-cell pipeline which stores event data and a 4-channel, 128-cell deep, analogue sampler which can be set to sample up to a frequency of 1.1 GHz. The chip operation is governed by 75 parameters[7] representing 239 bits which are spread in several registers set by slow-control.

The standard running mode is as follows. When the signal crosses the threshold of the comparator a level 0 (L0) trigger pulse is sent to the DAQ board. An external level 1 (L1) trigger from coincidences between OMs can also be requested. At the same time, if the discriminating (PSD) mode is enabled, the PSD block analyzes the shape of the signal and compares it to a predefined template. In parallel the pulse is sampled and integrated. The arrival time is given by a time stamp (TS), based on the 20 MHz synchronization clock, for coarse measurement and by a TVC value, which provides a fine measurement with sub-nanosecond resolution. Both values are measured when the PMT signal crosses the comparator threshold.

Three DACs control the three threshold comparators, one for L0 and two for the PSD[8]. In order to make the thresholds independent of the baseline, current DACs are used. For each threshold an external resistor connected between the signal input and the second input of a comparator convert the current into voltage. To this voltage the baseline voltage is added, as shown in Figure 7, so the comparison is baseline-independent.

[7] A part of the parameters control the response of the chip: thresholds, gate durations, sampler gain, internal data writing and resetting durations, ADC gain and offsets. Other parameters enable functions (PSD or SPE-only etc.), define bias currents, frequencies (submultiples of the clock). Some of the parameters add test functionalities for debugging or disabling some of the functions. For example, one may set a bit to trigger all events without waiting for an external signal. Parameters also allow testing the data readout machine, sending out pseudo-events to check the format of the data or programming the configuration of a special pin to spy on various critical internal signals.

[8] One threshold defines the minimum amplitude of the template. The other, higher, triggers the recording of the low gain channels: anode/5 and dynode 12.

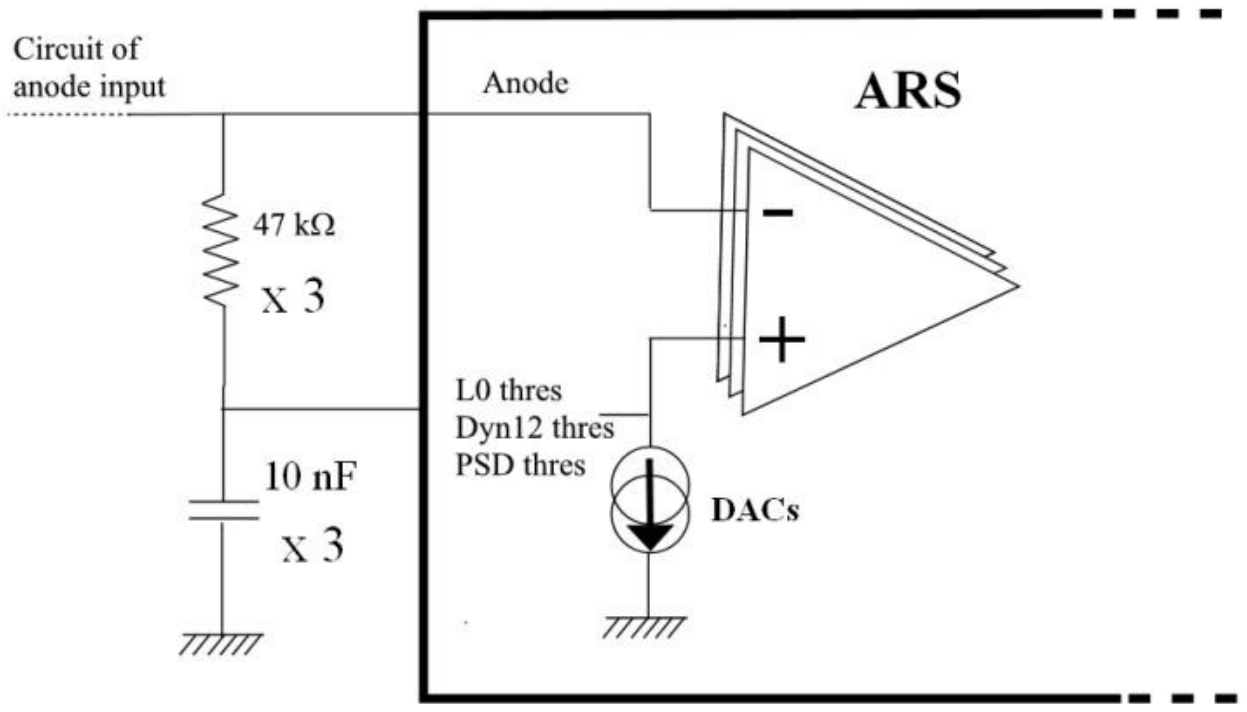

Figure 7. Circuits of the three threshold comparators.

The pulse charge is integrated by parts. A set of three switched capacitors samples the anode signal in a cycle of a period longer than the rise time of the pulse, typically 8 ns. At any given time one capacitor is integrating the current, another one is storing the charge of the previous period and the last one is being reset. When L0 occurs, the current integration is lengthened in order to encompass all the pulse charge; then the charge of the previous period is added.

At the end of the integration/discrimination gate the PSD returns a binary result: whether the pulse is of the SPE or WF type. In both cases this information is stored in the memory pipeline along with the pulse charge and the TVC and TS values. Cells for this pipeline are thus mixed: analogue (AVC, TVC) and digital (TS, PSD). If the selected mode is SPE mode, the only data that will be transmitted to the ADCs for digitization will be the pulse charge (the AVC produces a positive voltage proportional to the integrated charge) and the time (TVC). If the pulse is of WF type, sampling keeps running for the 128-sample depth duration before stopping. Signal samples are kept in memory in the fast sampling cells, outside the pipeline, awaiting a possible readout request. In the chosen architecture the ARS chip can store up to 16 SPE events but only one complex WF pulse in the sampler. Hence, the fast sampling is no longer available until the chip has finished processing the WF event, either to send data out or to clear the sampler cells. But as long as only SPE events occur the sampler remains available. The resulting dead times are discussed in section 6.5.

The analogue signals are processed by two identical 8-bit ADCs of successive approximation type. Conversion is clocked at a sub-multiple of the readout clock, 12.5 MHz, eight periods being needed to complete a conversion. Six analogue measurements with different dynamic ranges are digitized by only two ADCs; the least significant bit (LSB) value[9] and the maximum convertible voltage of each ADC (saturation voltage $V_{sat}$) are redefined for each event according to the channel being processed. These two parameters are selected among three possible pairs of values stored in register banks through slow control during the initial configuration of the chips. Each of the two provides a digital parameter value to a 5-bit DAC, defining the current for the LSB and the current for $V_{sat}$ respectively. The ADC response is therefore modified dynamically event by event, the readout sequencer selecting which of the 3 banks to use, depending on the data type to convert.

For the digitization of a stored event two options are available. Either all events are digitized without external trigger, or an L1 external trigger request signal must arrive within a predefined acceptance window associated with the event. The digitized charge, TVC and TS values, and, if available, the digitized WF samples are then sent out serially in a binary format. If the acceptance window ends without the occurrence of an L1 trigger, the pipeline cell associated with the acceptance window and the cells of the sampler are reset for use for a new event. The typical data acquisition conditions of ANTARES do not require offshore L1 triggers, so that all pulses above the L0 threshold are sent to shore.

Table 1 gives the circuit performance, where in particular it can be seen that the power consumption of the circuit is very low. An internal input buffer inserted between the anode input and the analogue memory to suppress the perturbation caused by the PSD response reduces the anode WF bandwidth from 130 MHz to 85 MHz.

[9] The voltage change needed to flip the LSB.

**Table 1**

ARS performance.

| Parameters | Measured values |
|---|---|
| Power consumption | 190 mW (at 5V power supply) |
| Sampler frequency | 300 MHz to 1.1 GHz |
| Sampler noise (RMS) | 5 mV |
| Dynamic range of each channel | 4 V |
| "Waveform" mode gain | 0.9 V/V |
| Integrator dynamic range | 130 pC |
| Integrator transfer function | 4.4 mV/pC |
| Integrator integral linearity | 1% |
| Input bandwidth | 130 MHz (85 MHz with buffer) |
| Readout clock | 10 MHz to 25 MHz |
| TVC transfer function | 44.5 mV/ns |
| TVC noise (RMS) | 800 ps |
| TVC integral linearity | 200 ps |
| ADC | 8 bits (6 bits effective) |
| ADC dynamic range | 1 V to 4 V |
| Max hit rate | 500 kHz |

## 4.2 Token ring communication

A token ring protocol is used to chain the two ARS chips serving the same OM, in order to decrease the acquisition dead time. Each ARS includes a register named "token". If the token is set to 1, the ARS owns the token and will treat the incoming events. Otherwise it will ignore them. At any time one and only one ARS should own the token. The token is passed from one ARS to the other with a delay of about 10-20 ns after the end of the integration gate. Each ARS presents three differential input pairs and three differential output pairs which connect the two chained ARSs and pass the token with a protocol depending on the availability of the chips and on their status of event processing.

## 4.3 Physics specifications

To meet the physics specifications the requirements for the ARS chip are: measurement of the arrival time of the anode signal with a precision better than 1 ns and of its charge to better than 10%.

The angular resolution of the reconstructed particle tracks depends on time precision, which is the main constraint placed on electronics. ANTARES is designed to have an angular resolution of less than 0.3° for neutrino energies in excess of 10 TeV, which relies on good positioning accuracy and good timing resolution of the signals recorded by the OMs. The specification for the timing resolution is such that it should be limited by the transit time spread of the PMTs, which have on average $\sigma_{TT}$ = 1.3 ns, and by the effect of chromatic dispersion of the light in water, which is expected to contribute with a similar amount to the time uncertainty[10]. To achieve this specification, all electronics and calibration systems are required to contribute less than 0.5 ns to the overall timing resolution. The front-end electronics precision depends on the TVC measurement precision, on the error on the walk correction (related to the charge measurement precision) and on the L0 threshold.

The absolute timing is measured by a GPS receiver interfaced to the Master Clock server at the on-shore station. 1 ms accuracy on absolute time is enough for the study of transient sources. The main uncertainties in the absolute timing come from the time offsets and fluctuations in the path common to all the signals, i.e., the 40 km electro-optical cable between the junction box and the shore-station and are well below the requirements.

The energy of low energy muons contained in the detector can be measured from their range. For muons above 500 GeV producing large electromagnetic showers, energy can be inferred by the light emitted by the showers with a rough resolution, so the pulse amplitude precision requirement is loose and a dynamic range of 100 p.e. is considered sufficient. Nevertheless a tight requirement is needed on pulse amplitude precision due to the walk effect (influence of the pulse amplitude on the threshold crossing time).

The photoelectron peak value in the ADC range is chosen as a trade-off between precision on the amplitudes of single p.e. pulses and dynamic range. Precision is needed in event by event measurements and also in the measurement of the average p.e. spectrum. This has to be known well in order to:

[10] $\sigma$ = 1.5 ns at a distance of 40 m. Scattering would have a stronger effect [12], but direct photons are not concerned.

- correct the walk effect,
- monitor the PMT gains,
- have an absolute calibration using the average p.e. peak for energy and position reconstruction of muons and showers
- determine the L0 threshold value in p.e. units for reconstruction efficiency. The threshold has also to be precisely known in mV and
- determine the PSD thresholds.

### 4.4 Digital data formatting

The ARS buffers, de-randomizes the event flow, formats the events and serializes them toward the DAQ board. The DAQ board contains a high density FPGA in which the readout system of the chips is implemented [5]. The data from the chips are buffered by the FPGA in a 64MB SDRAM into separate time frames whose duration can be set to values between 10 and 100 ms. The readout sequentially selects the registers and the internal ADCs in order to provide their value at the output of the chip in the appropriate format. There are six different types of events read out by six different sequences: CRM, SPE, WF, WF&dynode, TS reset and Status. A Status event is sent by the ARS when it is enabled at the beginning of a run, it contains a header and the initial TS value. When concurrent events occur at the same time the operations are performed with a predefined readout priority. Each event type has an identification number (binary code). The sizes of the different event types are different. For the most frequent signals, the SPE, the size is 6 Bytes. Hence, using only SPE events allows increasing the transmitted data rate compared with full WF events which are 45 times longer.

The data of the ARS are clocked out at 50 MHz by the DAQ board (section 3) but due to the protocol, using every 2nd bit as a validation bit announcing to the FPGA that a data bit follows, the effective data transfer rate is 25 Mbits/s. The chip behaves like a data generator using the readout signal to directly write into some external memory. The two chips on the motherboard have some different functions. They can both generate SPE, WF, WF&dynode and TS Reset events, but only the ARS in the first position on the motherboard can generate CRM events and only the second can generate Status events. When events of more than one type are waiting for readout in a chip, the following order is used to format and output events: highest priority goes to CRM, then to SPE, WF, WF&dynode, then to TS Reset, and finally to Status. Events of the types SPE, WF, WF&dynode are separated from one another by inter-frame spaces of variable length during which TS Reset or Status events of lower priority can be processed even if the pipeline is not empty. Each event has an 8 bit header to identify the event format and the recorded mode of the event. The SPE event output duration is ~ 1.9 µs. The WF event output duration, including the anode WF, the clock sampling and nonvalid bits[11] is ~ 145 µs.

### 4.5 Other applications of the ARS

The circuit is an innovative development in microelectronics for physics because of the number of integrated electronic functions. It is actually a full system on chip which takes decisions online about how to process events, reducing the dead time.

The development of such a chip has led to several other applications. Two thousand chips of the preliminary version ARS0 equip the four cameras of the H.E.S.S. telescope [13]. A new chip, SAM, standing for Swift Analogue Memory, has been developed and equips the 2000 channels of the camera of the H.E.S.S. phase 2 mirror, currently under assembly [14]. Astroparticle physics can benefit from ASIC developments around front-end fast analogue sampling memories. Their versatility and reliability make them suited to most requirements [15].

From the acquired experience with ANTARES, a new ASIC is being designed for the future cubic-kilometre scale detector [16].

# 5 Production, selection and integration

Tests and integration involved several laboratories of the collaboration. The bare ARSs were tested, selected and calibrated in a laboratory by batches following production. Some tests and characterizations were also carried out after the ARSs were mounted on their motherboards and during the integration of the ARS motherboard in the electronics modules. Most of these measurements were meant as functionality tests. The parameters that are finally used for the data acquisition are those measured *in situ*.

[11] Compensating digitization delays.

The main selection of the chips was based on the first tests performed during production on batches of 330 ARS circuits, using functional and parametric tests, before the selected circuits were integrated on their motherboards. This selection procedure was quite effective, since a global analysis performed *a posteriori* on all the measurements performed on the bare chips and of integrated motherboards only led to the rejection of 4 faulty chips out of a total of 1900 integrated on motherboards for data taking.

The tests of the ARS circuits took place from September 2004 until October 2005. The total number of produced circuits is 6377, divided in batches. A pre-series of 200 circuits had been tested in 2003 to evaluate the yield and the number of circuits to manufacture.

During these tests, each circuit was tested individually with a set of 20 measurements. This was done on a test bench using FPGAs and Labview software. Because of the many functionalities and parameters of the chip, the tests were quite complex and not all the parameters were tested. The tests were done in two steps: the first series of tests checked the overall functionality of the circuit. The second series of parametric tests gave information concerning the performances of the circuit.

## 5.1 Functional tests

Functional tests were again sequenced in two steps:

- the ability of the chip to respond to slow-control and to deliver data was tested. These tests allowed to remove totally non-functional chips (e.g., due to a short circuit on the power supply).
- the chips were tested again for what concerns the following functionalities: LED pulser, CRM, time stamp, PSD in different configurations, token ring.

## 5.2 Parametric tests

The parametric tests were based on the measurement of parameters related to the chip operation, most of them depending on the 75 slow control settings (section 4.1). The chips were then sorted according to the value of these parameters.

Each time the measurements were finished on a batch of circuits, the data analysis software sorted circuits into three categories:

- **Data-taking** ARSs: circuits within specifications for the ANTARES data taking.
- **Control** ARSs: circuits good enough for the additional functions required on the ARS motherboard (third position).
- Rejected circuits.

Data-taking ARSs were selected to be implemented on the first and second position on the ARS motherboard, the control ARSs on the third position. The selection was done according to the following criteria:

- Fulfil the specifications and physics requirements. However, this did not place the most stringent constraints.
- Exclude atypical chips. For this purpose, the parameter value had to belong to the Gaussian distribution of the whole set, i.e., to be within $\pm 3\ \sigma$ of the mean value of the parameter for the full batch.

Data were stored in a database so that they could be used later to better understand the detector performance during data taking. The tested parameters were as follows:

- **Power consumption.**
- **DAC transfer functions:** Digital-to-Analogue Converters allow setting different thresholds on the PMT signal, from some mV to a few hundreds of mV. If the DAC response is not linear, it would be difficult to control the data stream.
- **Charge integration gate**. The gate is set typically around 25 ns after threshold crossing plus 8 ns before. With a common setting for all chips the spread is 1.5 ns.
- **Integrator cycle period** (=additional integration gate before threshold crossing).
- **WF signal:** transfer function of the sampler (gain and offset).
- **Pipeline writing and resetting duration.** It plays a role in the dead time of the SPE processing.
- **L1 gate and delay.**
- More parameters: **integration gate**, parameters concerning the measurements of **time** and **charge**, the **L0 trigger threshold** and the **ADCs**. Tests on these important parameters are presented in more detail in section 5.3.

## 5.3 Results of tests

During the functional tests it was found that an incorrect TS value is sometimes recorded when the rate of input signals exceeds 100 kHz. The principle of an asynchronous design is that each signal can change the state of the system inside the chip, but at very short time differences between signals (less than 100 ps), as may happen with high signal rate, some state changes may not be correctly processed. The probability of a TS being affected by these bit-flip problems is between $10^{-5}$ and $10^{-3}$ depending on the input rate, which does not affect the apparatus performance.

Concerning the parametric tests shown below they come from analyses performed on a subset of the 3070 functional data-taking circuits, including all the circuits mounted on the detector and a part of the spare ones.

### *5.3.1 TVC parameters*

The ARS has two TVC ramps, shifted in time, to avoid uncertainty on the time measurement when the ramp goes down to zero. Time measurements are performed alternatively with ramps 0 and 1. The transfer functions of the two ramps may not be identical. Hence, an important parameter of the ARS is the difference between them: if the slopes do not differ by more than 5%, one can use a unique value per ARS to reconstruct the event time. Consequently, circuits with this feature were selected.

Figure 8 shows the distribution of the slopes of the TVC ramp 0 for the full set of the 3070 data-taking ARSs. The slopes of ramp 1 follow a similar distribution. The difference of average slopes is less than 2%, see Table 2.

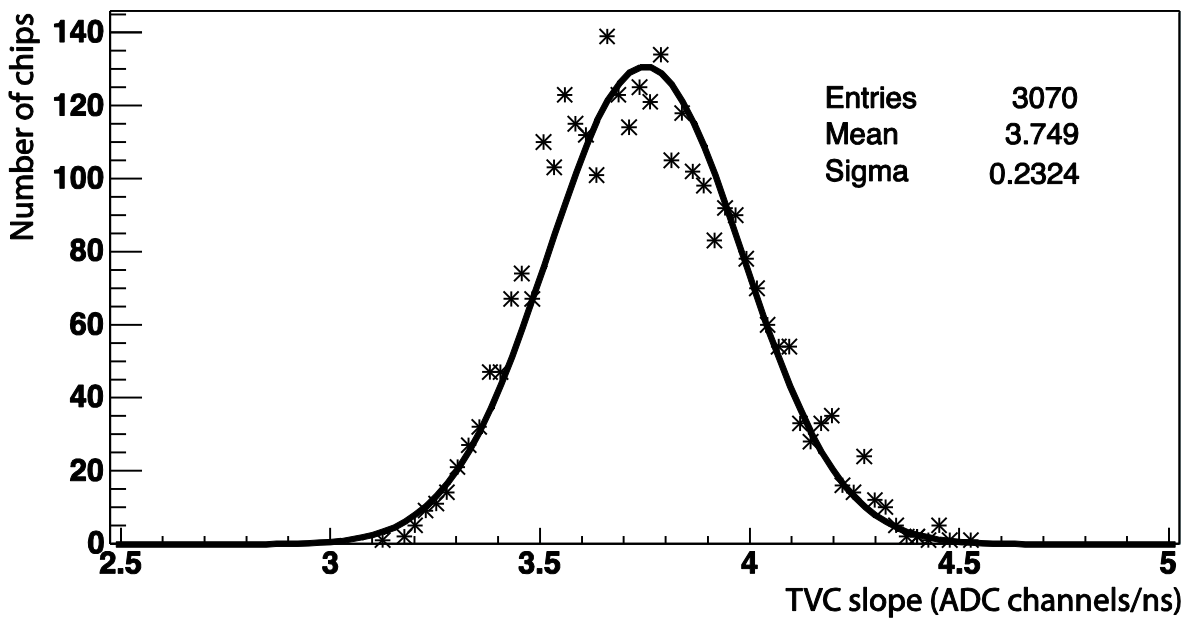

Figure 8. Distribution of the TVC slopes of ramp 0 for the data taking ARS's. The Gaussian curve superimposed to the data is only indicative.

Table 2

Main measured chip parameters.

| | Mean | RMS |
|---|---|---|
| TVC ramp 0 slope [*bits/ns*] | 3.75 | 0.23 |
| TVC ramp 1 slope [*bits/ns*] | 3.82 | 0.24 |
| TVC noise [*ns*] | 0.45 | 0.016 |
| AVC transfer slope [*bits/pC*] | 1.218 | 0.055 |
| Threshold transfer function intercept [*DAC units*] | 14.74 | 7.8 |
| Threshold transfer function slope [*mV/DAC unit*] | 1.075 | 0.1 |

A second important parameter is the RMS noise on the TVC measurement because it affects directly the precision of the track reconstruction. The electronics must minimize this error in order not to influence the total error of the system. With a noise around 450 ps on the average (Table 2), the selected circuits fulfil this constraint.

### *5.3.2 AVC parameters*

The transfer functions of the charge measurement through the integrator and AVC were measured, including the noise and linearity values. The distribution of the transfer function slopes is close to a Gaussian with a σ of 0.055 bit/pC for a mean value of 1.218 bit/pC for the reference ADC setting. This relative dispersion of about 5% at one σ shows that the integrator-AVC component does not produce significant gain non-uniformities. Hence, only the PMT and ADC gains will have to be considered for gain equalization of the different OMs.

### *5.3.3 The effective threshold parameters*

**Raw threshold:** The output voltages for the three 8-bit threshold DACs (L0, PSD, Dynode 12) were measured by a voltmeter. The transfer function of each of these DACs is linear, so for each DAC, a slope and an offset are extracted. By construction, the offset is always 0, so the exclusion criteria are only set on the slopes. The ±3 σ selection here means roughly ± 15 % of the mean value. The observed spread is mainly due to the spread of the chip physical parameter setting the current, i.e. the threshold voltage $V_t$ of the CMOS transistor and to cumulated mismatch on 4 levels of current mirrors inside the chip.

The voltage measured with the method above is not the real threshold value. It must be convoluted with the discriminator transfer function, limited by the bandwidth, and with the discriminator offset. So the relevant parameter is the effective threshold.

**Effective threshold**: The efficiency curve was measured for input signals of 60 and 90 mV as can be seen in Figure 9. The rate is monitored by the CRM events varying the L0 threshold settings. Plotting the CRM mean for different threshold values the usual S-curve, feature of discrimination, is obtained. The threshold value at 50% of the CRM maximum is the threshold at which half of the pulses trigger. For each ARS, from the two curves with inputs of 60 and 90 mV, by linear interpolation, the intercept and the slope of the effective L0 threshold transfer function are extracted. A large threshold spread, around 40 DAC units, is found. As shown in Table 2, the offset spread corresponds to 6 mV RMS (7.8 DAC units of 750 µV each), which is compatible with the expected offset spread of an uncompensated CMOS input comparator.

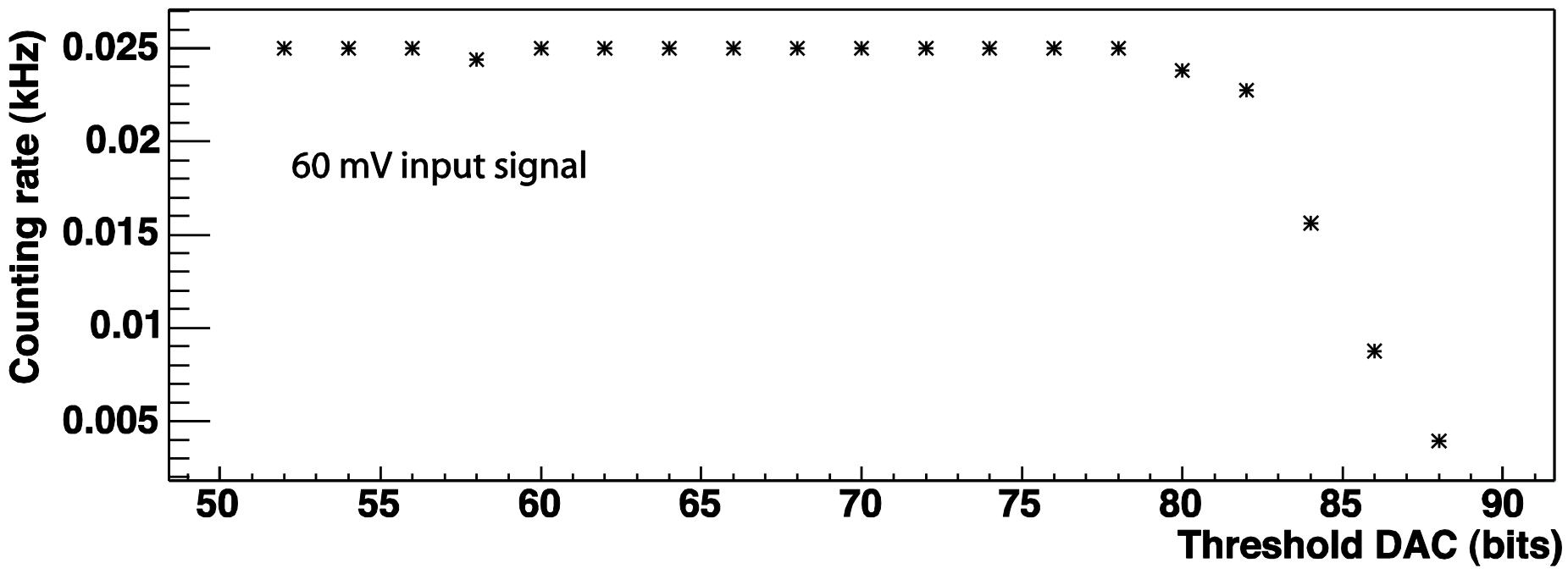

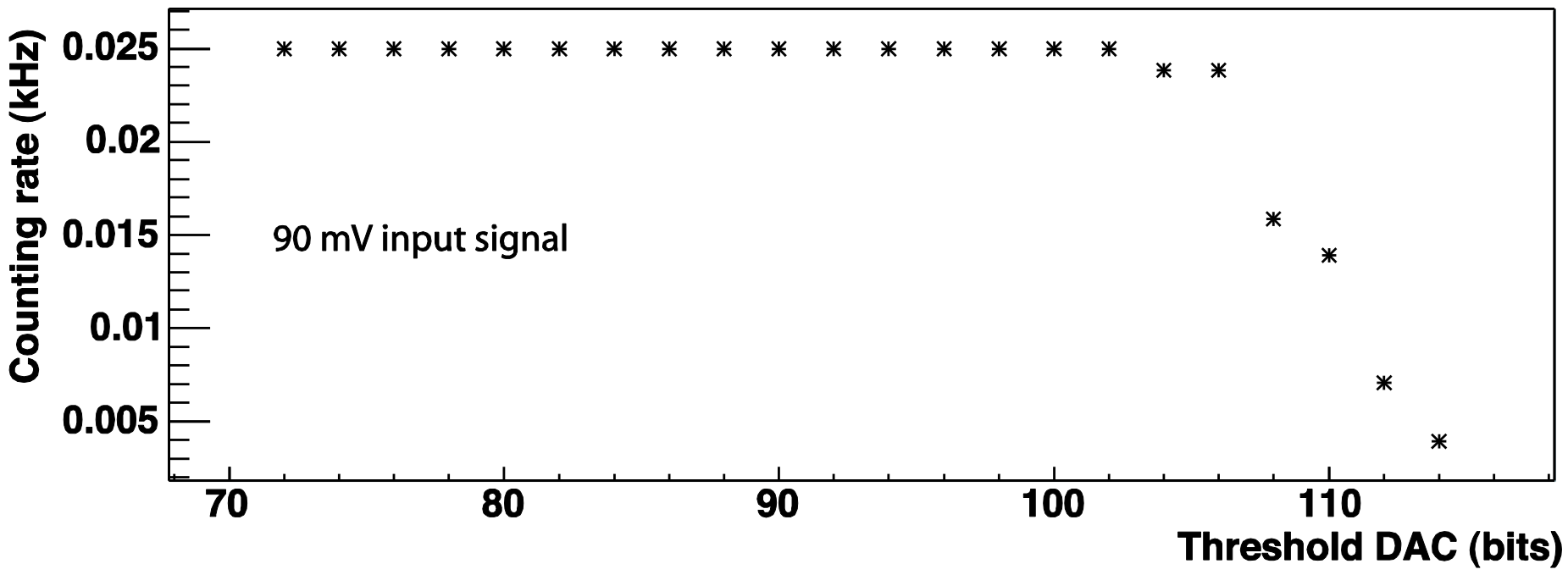

Figure 9. Efficiency curves for L0 threshold with input signals of 60 mV and 90 mV. Interpolation between the two measurements gives the response slope.

#### *5.3.4 ADCs*

As mentioned in section 4.1 the ADC transfer function is governed by two parameters, the LSB and the saturation voltage $V_{sat}$. The two ADCs of each chip were tested with DC signals over their entire dynamic range through the debug input, for two different slow-control settings. For each setting and each ADC the distribution for the maximum value $V_{sat}$ , the minimum value and the LSB were obtained. The tests showed that:

- the behaviour of the two ADCs is very similar.
- the ADCs have a satisfactory integral linearity for input levels between 1 and 4.9 V. A differential non-linearity effect will be discussed in section 6.1.1.
- there is a very large spread of $V_{sat}$, attributed to the DAC controlling this voltage. This spread increases with the voltage value. This spread is 400 mV peak to peak for a $V_{sat}$ = 2.2V. This results in a spread of the pedestals and dynamic ranges which can be corrected by tuning the settings.
- the spread of the LSB increases with decreasing LSB. It reaches ±10% for small LSB.

## 5.4 Yield

15 batches of 330 chips, out of the 19 foreseen in total, were sufficient to provide the whole set of circuits needed for the construction of ANTARES. To build the 12 detection lines, the collaboration needed about 1800 ARSs in positions 1 and 2 (used for data acquisition) and 900 in position 3 (using only the DAC function) on the motherboard. Figure 10 shows the yield from batches in chronological order. The global yield is more than 51%, lower than usual, because of the complexity and multiplicity of functions of the circuit. A sufficient number of circuits are available to maintain the experiment during 10 years with the possibility of a 5% replacement per year.

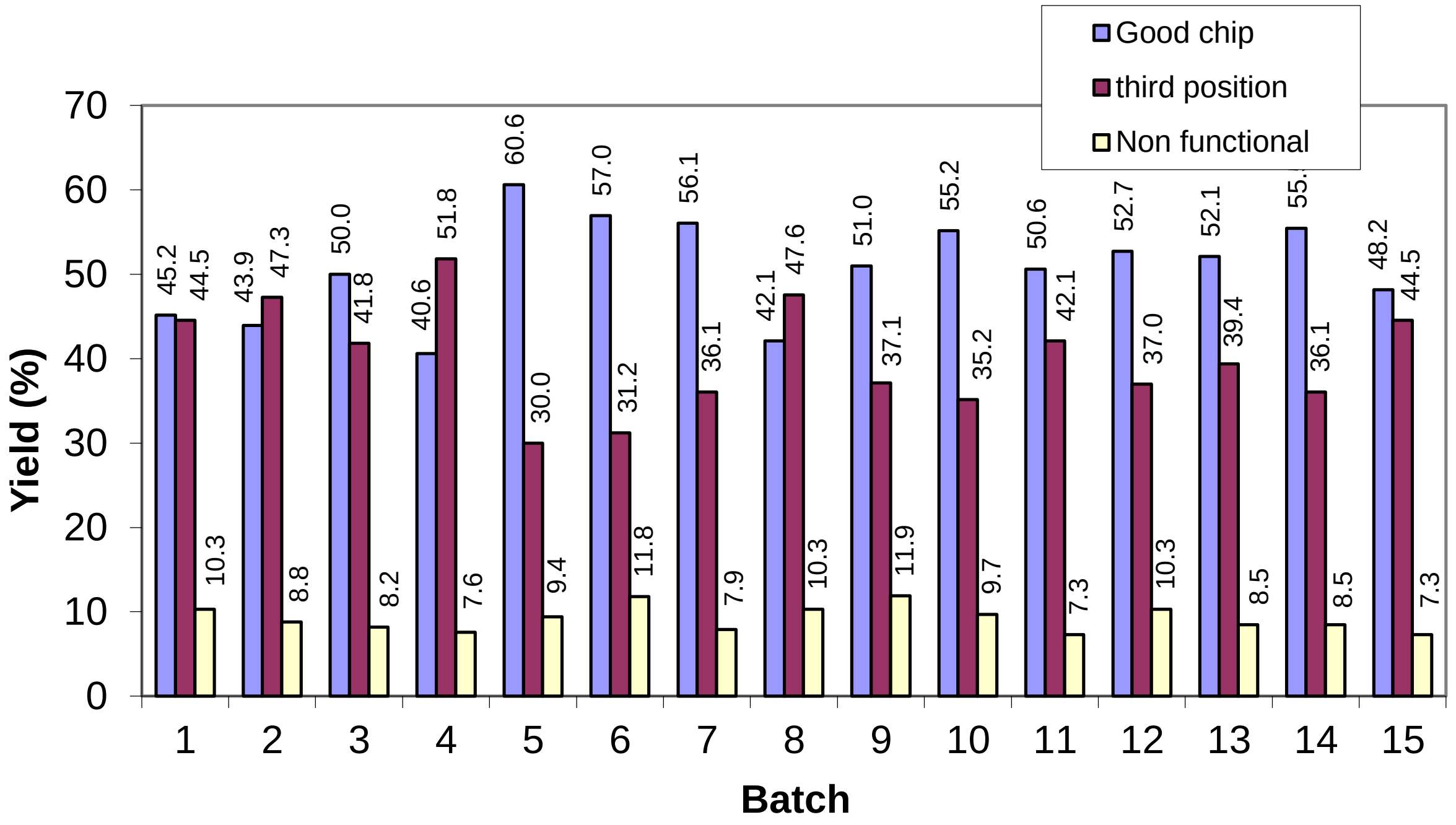

Figure 10. Yields from tested ARS batches. Out of 4822 chips 2445(51%) were found good, 1938 (40%) to be used for control only and 439 (9%) non functional.

# 6 Performances *in situ*

## 6.1 Charge measurement performance and calibration of the photoelectron response

The L0 threshold is set typically to 1/3 p.e. The dynamic range of the AVC and anode WF channels is between 20 and 30 p.e. (for an LSB of ~ 7 mV), while the low gain channels allow the sampling pulses of up to 800 p.e. amplitude. The conversion of ADC counts into p.e. units, which is needed for the physics analysis, is done assuming a linear response of the integrator and the ADC. Special runs reading the PMT current at random times[12] allow the measurement of the corresponding pedestal value of the AVC. In addition the p.e. peak can easily be studied with minimum bias events since the optical activity due to $^{40}$K decays and bioluminescent bacteria typically produces single photons. The knowledge of the p.e. peak and the pedestal is used to estimate the charge over the full dynamical range of the ADC. As shown in Figure 11 the hypothesis of a linear response of the AVC has been verified in the laboratory sending signals of a pulse generator to a standard LCM.

### *6.1.1 ADC differential non-linearity (DNL)*

The integral linearity of the ADC used in the ARS chip has been independently studied using the TVC and is satisfactory. But the ADCs suffer from differential non-linearities, i.e., inhomogeneous bin sizes. The simplest ADC design was chosen in order to minimize power dissipation and the size of the chip. The 8-bit design is in fact effectively 6-bit, conforming to the original physics specifications which did not focus on a need of a high charge resolution.

An example of *in situ* charge distribution on raw data, mainly bioluminescence and $^{40}$K background, is shown in Figure 12. The methods used to estimate the pedestal and the p.e. peak treat integrated distributions in order to reduce the systematic effects from the DNL.

[12] The slow control triggers the ARS readout independently of any signal at the input. The HV is not turned off. The probability that an optical background event falls within the integration gate is negligible, most of the time the signal is null.

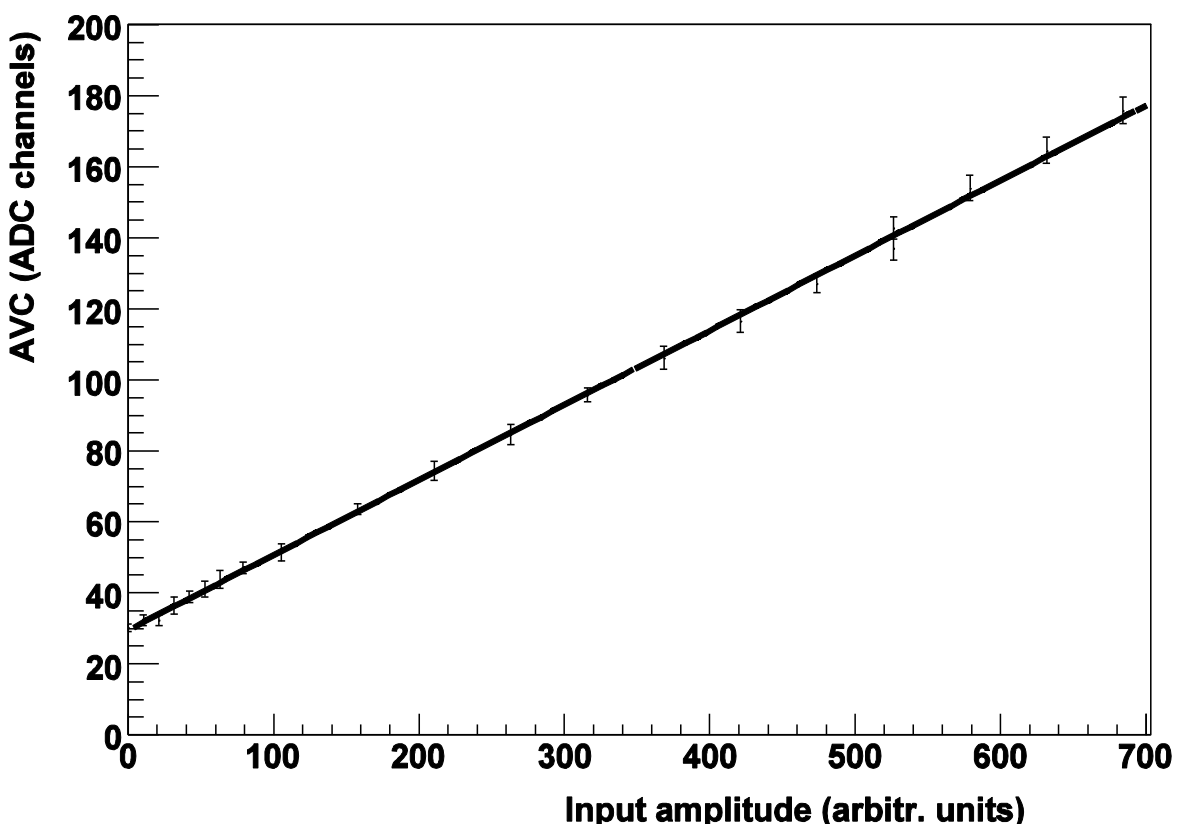

Figure 11. Charge measured in the AVC as a function of the injected pulse amplitude in laboratory tests.

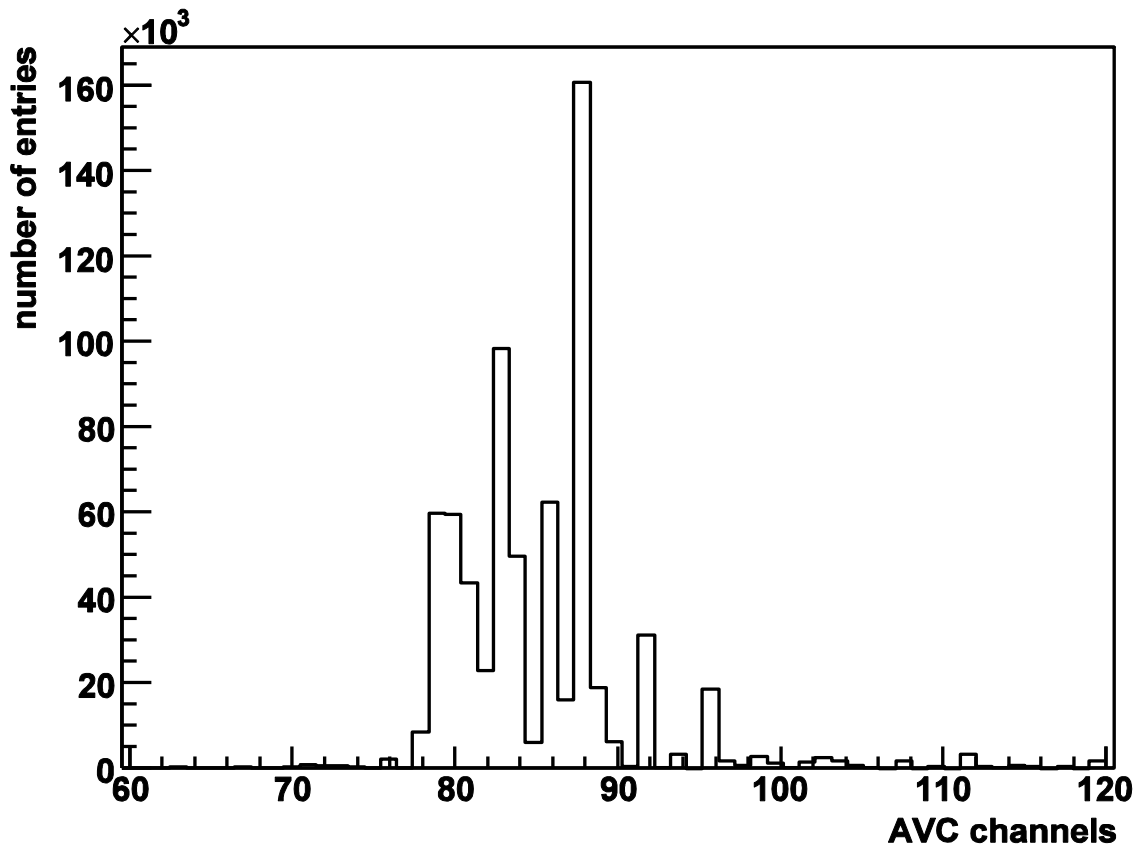

Figure 12. Example of AVC distribution of p.e. data. The observed structures, especially empty channels, are due to the DNL of the ADC.

### *6.1.2 Pedestals*

During pedestal runs the mean AVC value corresponding to a null signal in the ARS is measured *in situ*. The AVC distributions obtained are integrated to smooth the DNL and finally fitted with an Error function. Extracted pedestal values from *in situ* runs have been stable in time since the immersion of the first line. The comparison between the pedestal as measured *in situ* and the intercept at null charge measured during electronics integration shows a good agreement.

### *6.1.3 Charge gain*

The difference in ADC channels between the one p.e. peak and the pedestal is used to convert individual measurements into p.e. units. Charge distributions obtained with all hits above L0 in a PMT are parameterized with the following formula:

$$dN/dx = Ae^{-\alpha(x-x_{th})} + Be^{\frac{-(x-x_{pe})^2}{2\sigma^2}}, \qquad (1)$$

where $x$ is the charge in ADC units, the first term accounts for the dark current of the PMT and the second one approximates the p.e. distribution. The parameters $x_{th}$ and $x_{pe}$ are respectively the effective threshold and the p.e. peak in ADC units. A primitive of the function above is used to fit the cumulative distributions.

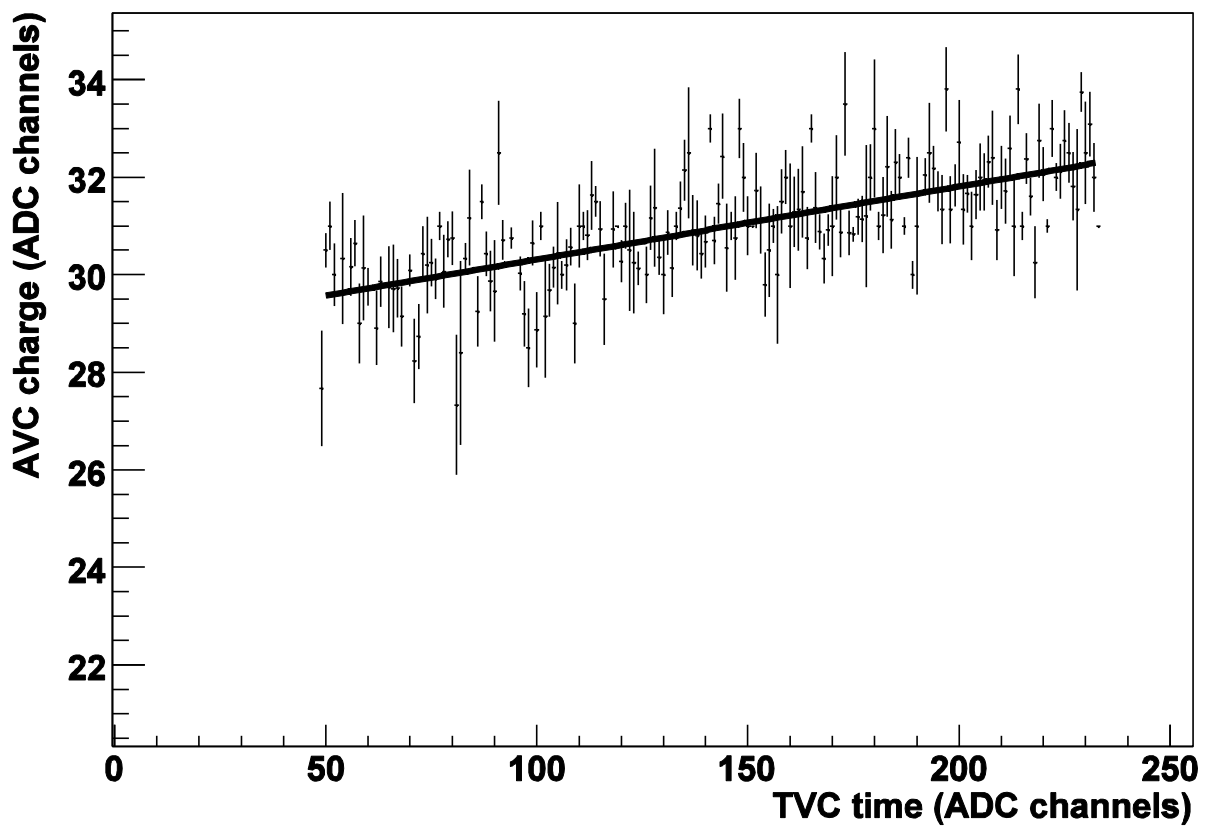

Figure13. Example of the observed cross talk effect affecting the charge measurement.

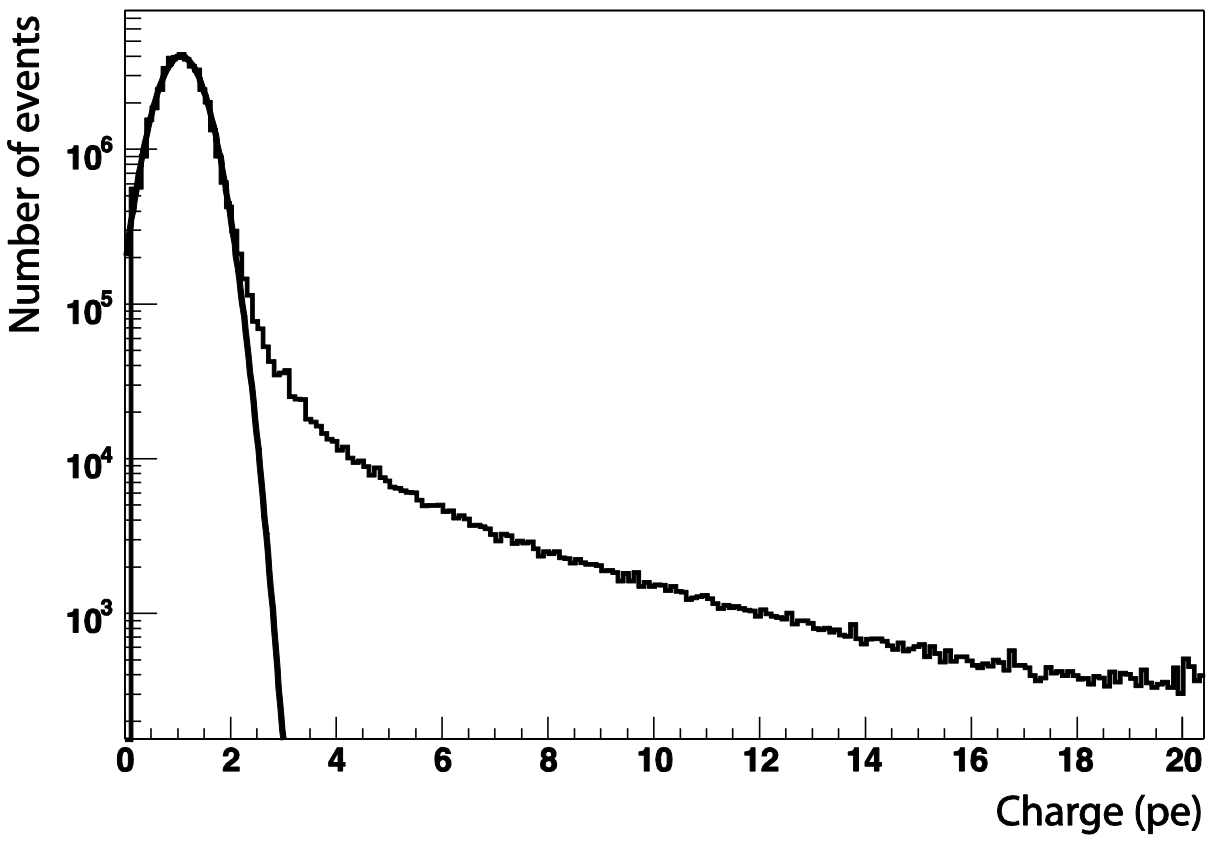

Figure 14. Example of overall charge distribution in p.e. units obtained with 10 lines after calibration. The Gaussian fit of the p.e. peak gives a mean value of 1.05 p.e. with a σ of 0.4.

### *6.1.4 Time and charge cross-talk effect*

The charge measurements in the AVC are affected by the time measurements in the TVC. The inverse effect does not occur. This effect is thought to be a cross talk of the capacitors inside the ARS pipeline and can be as high as 0.2 p.e. on an event-by-event basis. It is a linear effect that does not require correction on a high statistics basis; when hits populate the full range of the TVC the effect washes out. Nevertheless a correction should be applied to the measured charge of a single event. This correction can be inferred with *in situ* measurements by plotting the AVC value against the TVC value as can be seen in Figure 13.

A distribution of charges in all PMTs from two physics runs is presented in Figure 14 after the full calibration procedure presented in this section. Most of the events are in the p.e. peak. Higher charges are due to $^{40}K$ decays in the glass sphere and in the water, to bioluminescence bursts and to atmospheric muons. The main systematic error of the charge measurement is due to the measurements of the pedestal and of the p.e. peak because of the DNL. The accuracy of the measurement is estimated to be of the same order as the r.m.s. of the distribution of the PMT response to 1 p.e., which is ~0.3 p.e.

## 6.2 Thresholds: *in situ* measurement and equalization

When the time over threshold of a pulse is too short, i.e. when the hit amplitude is just above threshold, the ARS chip cannot properly generate the time stamp of the event, which remains null. However, the charge is recorded correctly. This specific behaviour has negligible influence on efficiency but enables the measurement of the effective threshold in AVC units[13] by selecting events with null time stamp. During special calibration runs thresholds are varied by slow control and the mean AVC values of events at the threshold are recorded. The result of the linear fit of the transfer function gives the intercept (DAC value for null threshold) and slope. An example is shown in Figure 15. This method can be applied to every data-taking ARS of the detector, yielding individual *in situ* calibration and transfer functions, which are stored in a dedicated database and used to adjust the thresholds.

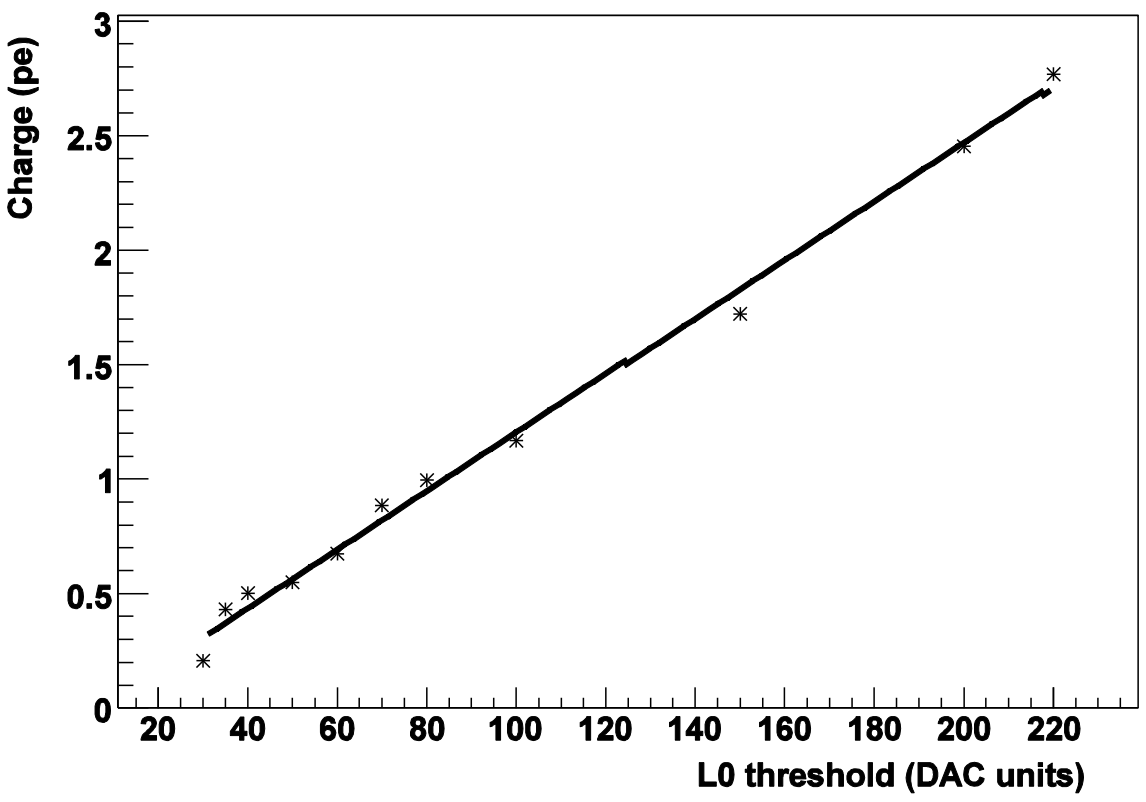

Figure 15. Effective transfer function of the threshold DAC in units of p.e. of an ARS. The local deviations from the linear fit are due to the ADC differential non-linearities.

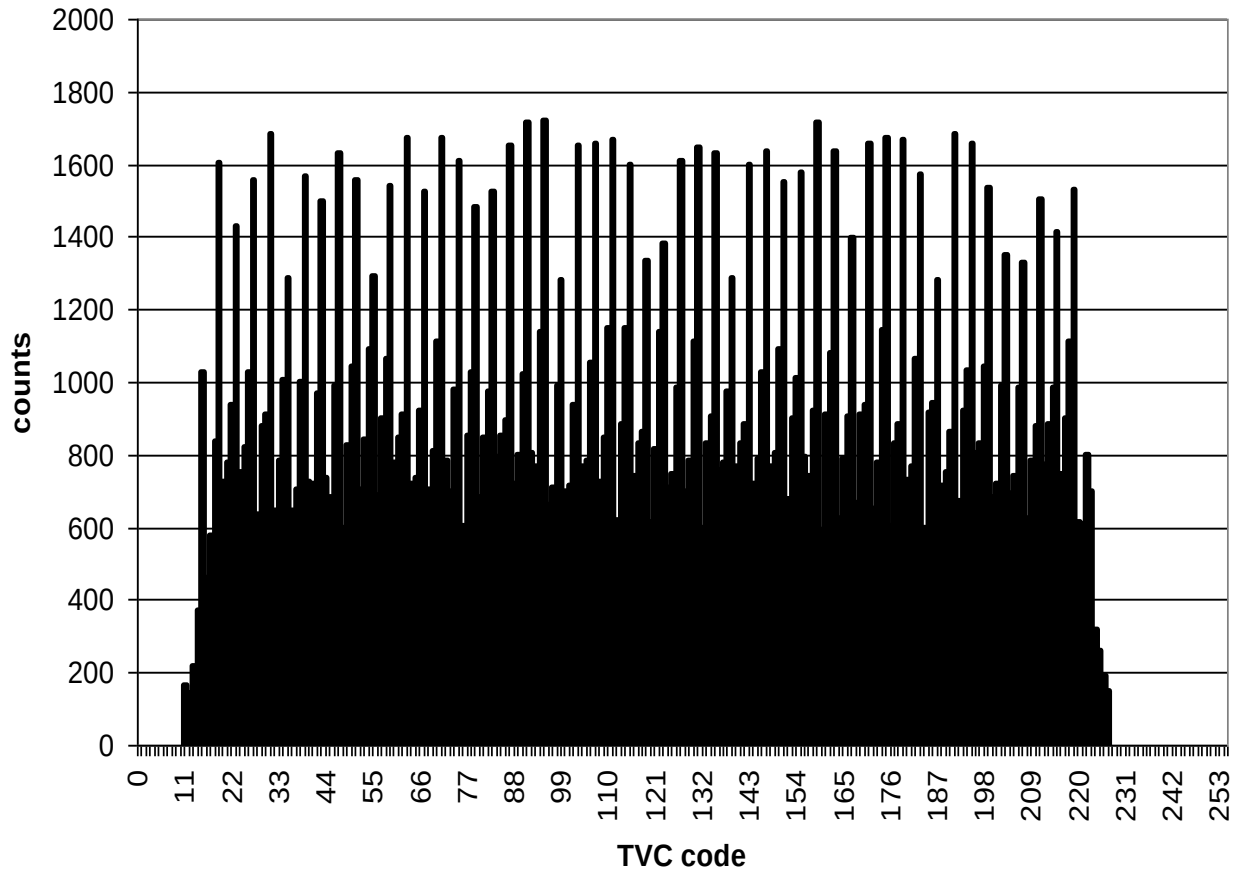

Figure 16. The flat time distribution of random events over 50 ns as measured in the ADC of the TVC.

Thresholds are found to be stable in time in absolute value (mV), but they have to be adjusted occasionally in p.e. units in order to match the gain drift of the PMTs. The result of the tuning is checked looking at the counting rates of all detector channels. Because of the varying optical activity in the sea water, the absolute rates are irrelevant for that matter, but a uniform response of the detector channels is expected, when integrated over a few minutes. We therefore check that the new settings indeed reduce the rate dispersion. This can be achieved in an almost unbiased way using data without on-shore trigger. Threshold equalization improves the rate dispersion typically from 15% to 10%. Looking at the two ordered time differences between consecutive hits on chips of the same OM, one can even determine threshold differences within the pair even though their rates are equal (since the two chips are in flip-flop, they trigger in turns and their rates are equal, imposed by the chip with the higher threshold).

Threshold differences within the pair of chips of the same OM are computed using the two distributions of the ordered differences between consecutive hits, $t(ARS_1)$- $t(ARS_2)$ and $t(ARS_2)$- $t(ARS_1)$. As expected from Poisson statistics, these two time difference distributions are exponential, each with an exponent directly linked to the effective rate of the single ARS.

[13] The AVC value, integrated pulse, is approximately proportional to the amplitude maximum.

## 6.3 Time measurement precision

Time precision is a crucial requirement for the experiment and the strictest specification for the electronics, as pointed out in section 4.3.

The raw TVC data show the typical differential non-linearity due to the ADC as is shown in Figure 16. As a consequence, the ADC is a 6 effective bits instead of the designed 8 bits but the error remains below 0.5 ns.

The time calibration of the detector is performed using several systems which provide complementary information on the propagation of the signal in the detector, namely a 20 MHz reference clock calibration system, LEDs installed inside each OM and the Optical Beacons, equipped with LED or laser devices. Four LED Beacons are distributed along every line [17]. Each beacon contains 36 individual blue LEDs (dominant wavelength 470 nm) synchronized in time and arranged to give a quasi-isotropic light emission. A small PMT internal to the LED beacon monitors the output light pulse timing and amplitude and provides the time reference of the light flash. The time resolution of the electronics can be deduced from the time difference distributions between the emission from the LED beacon on a storey of a given line and the arrival of the light at the OMs of the next storey in the same line. Due to the short distance between the beacon and the OMs of the closest storey (13.4 m) and the high light intensity of the beacon pulse, contributions from the transit time spread in the PMT, from the signal pulse walk, as well as from light scattering and line movements are negligible. Hence, the width of the measured time distribution, fitted with a Gaussian of typically $\sigma \sim 0.4$ ns reflects the resolution of the readout electronics and is in agreement with the expected time resolution [6].

The time precision has been cross-checked using $^{40}$K coincidences [18].

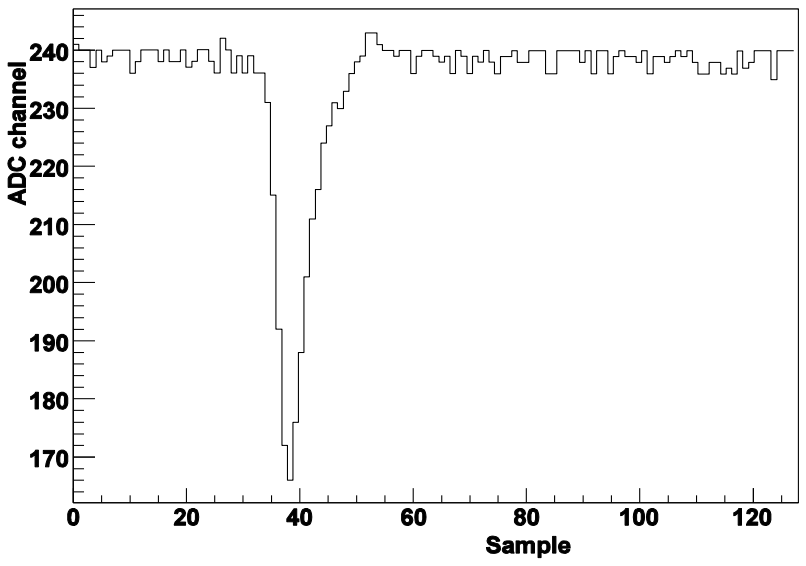

a. Single pulse.

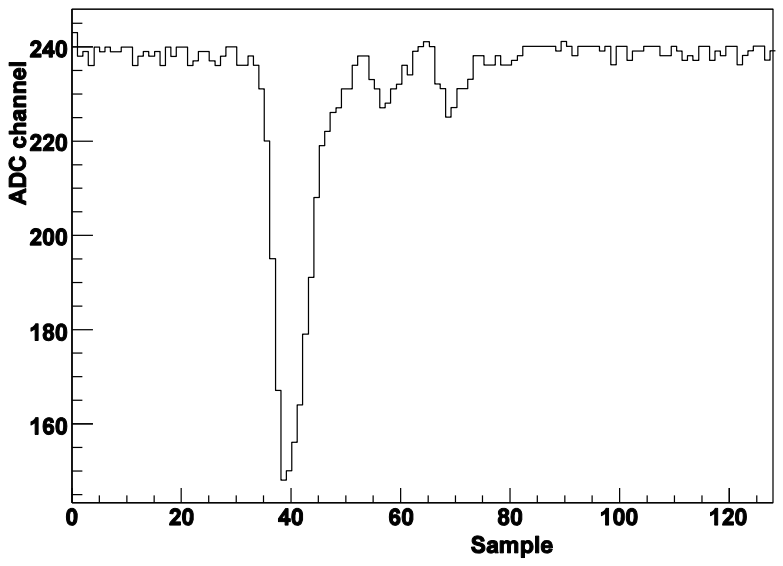

b . A pulse with PMT after-pulses.

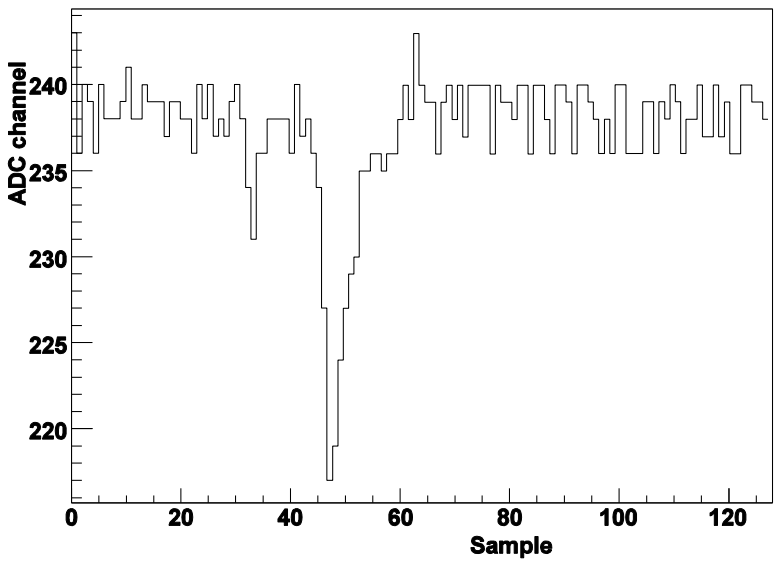

c. Two pulses arriving in the same time window.

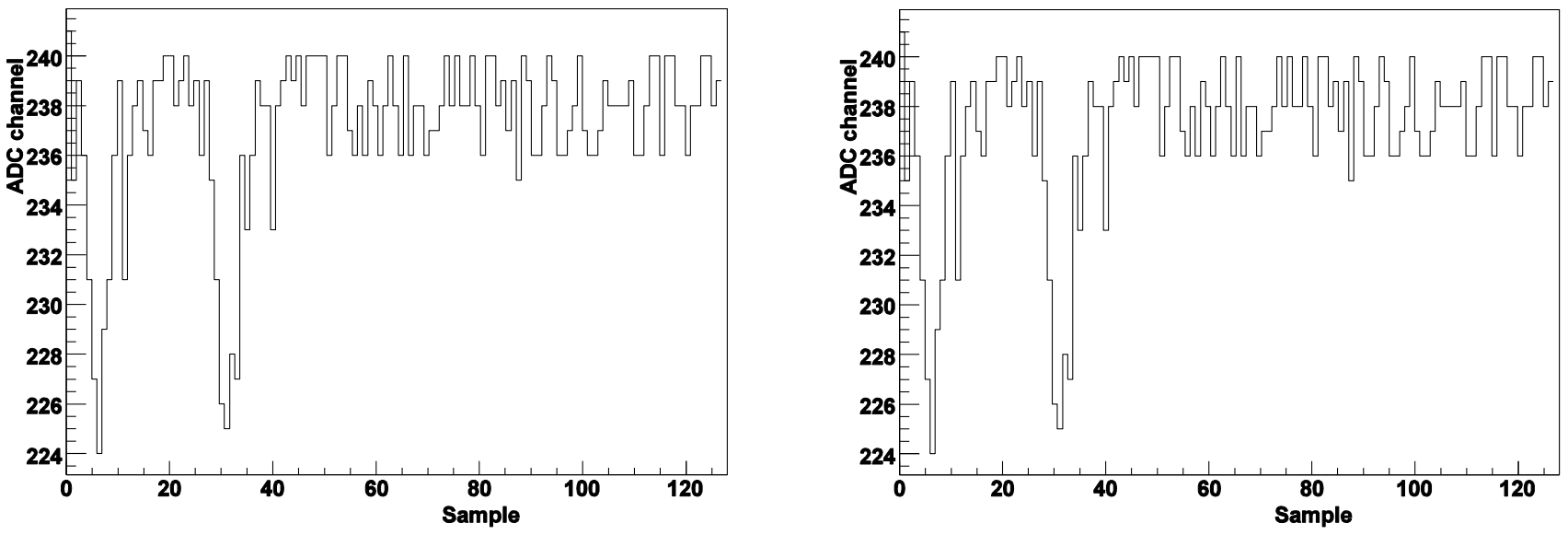

d. A sequence of pulses recorded by the two ARSs of one OM.

Figure 17. Waveform events with 640 MHz sampling.

## 6.4 Waveform events

Using the waveform (WF) mode the PMT signal can be sampled and recorded in detail. Figure 17 shows typical signals observed in the sea (bioluminescence and $^{40}$K decay Cherenkov light). The resolution of the WF sampler is good enough to distinguish between two consecutive pulses. The noise is about 5 mV (RMS) for the chosen sampler speed of 640 MHz. The system is able to record up to 256 cells thanks to the token ring mechanism (see Figure 17d). Figure 18 shows the distribution of integrated WFs recorded in an ARS from a sample of events (mainly single p.e. events from bioluminescence and $^{40}$K).

So far, the WF has been used for calibration purposes, in dedicated acquisition runs. The walk correction calibration constants are obtained from these runs and are used to correct the time of each hit during track reconstruction. Figure 19 shows the correlation between the charge measured in the AVC and the sum of the sample cells for WF events.

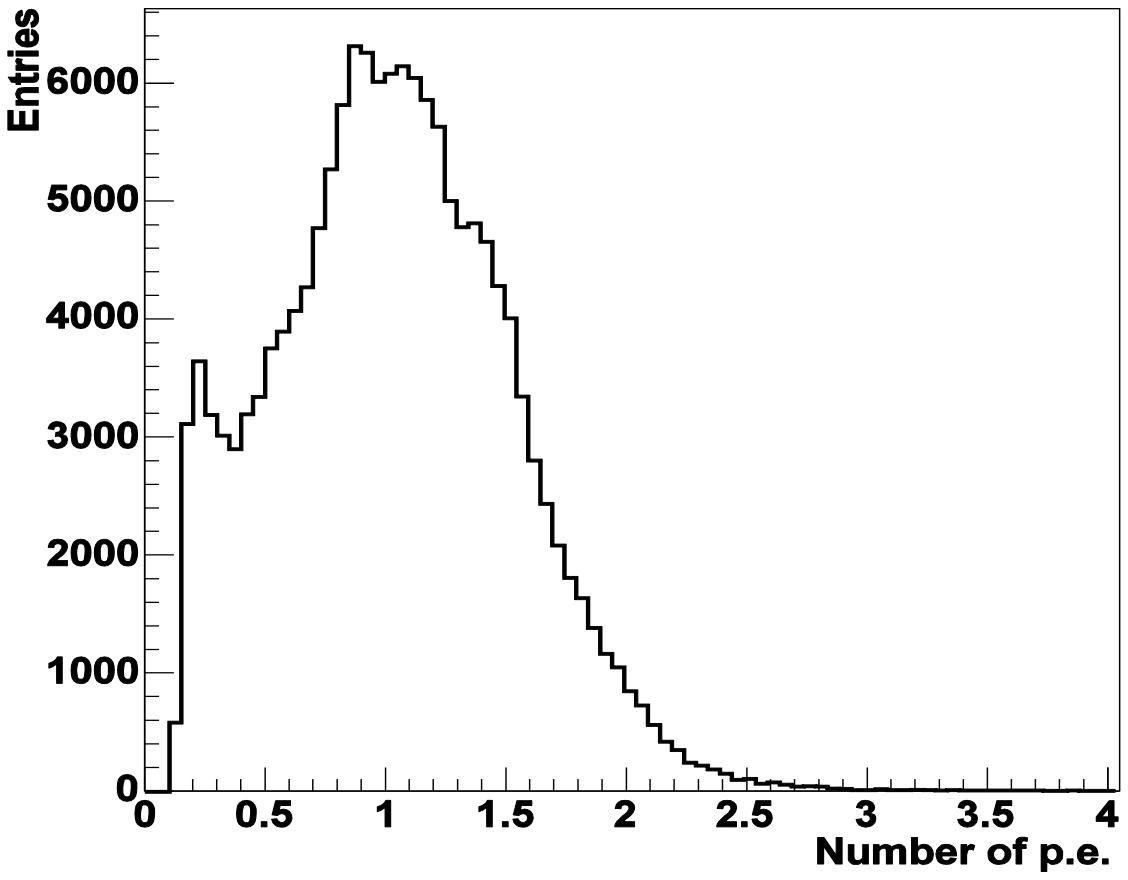

Figure 18. Distribution of integrated waveforms recorded by one ARS.

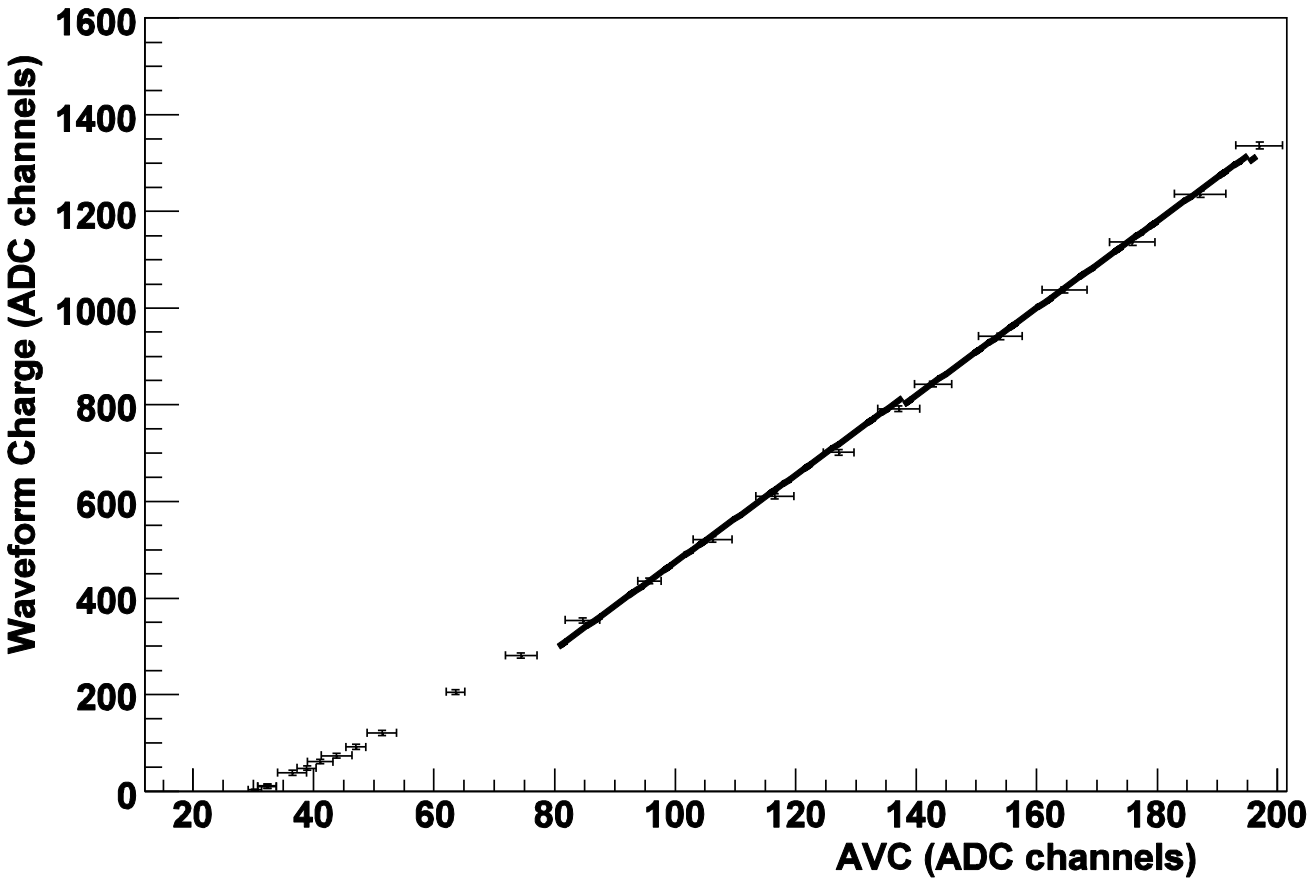

Figure 19. Correlation between the charge measured by the AVC and the sum of the waveform sampling cells for WF events. It is linear for charges above 5 p.e.

## 6.5 Dead time

The ARS anode input bandwidth in sampling (WF) mode is 85 MHz, so consecutive pulses can be separated down to 5-10 ns. The maximum frequency of L0 trigger pulses that can be counted by the FPGA on the DAQ card is 9 MHz.

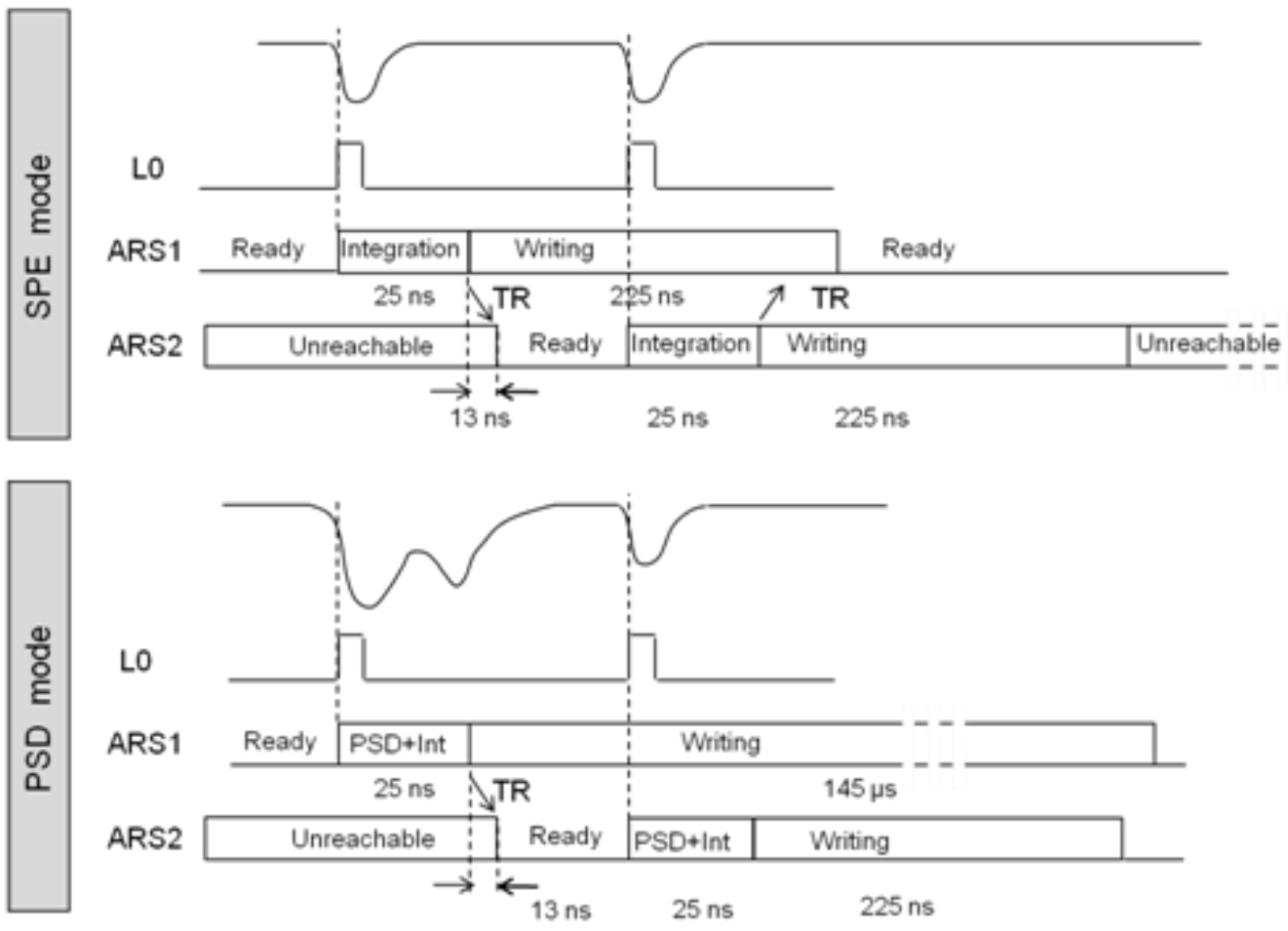

Figure 20. Timeline of two consecutive events and the corresponding occupancy in the front-end for the cases of SPE only and mixed (PSD) operation.

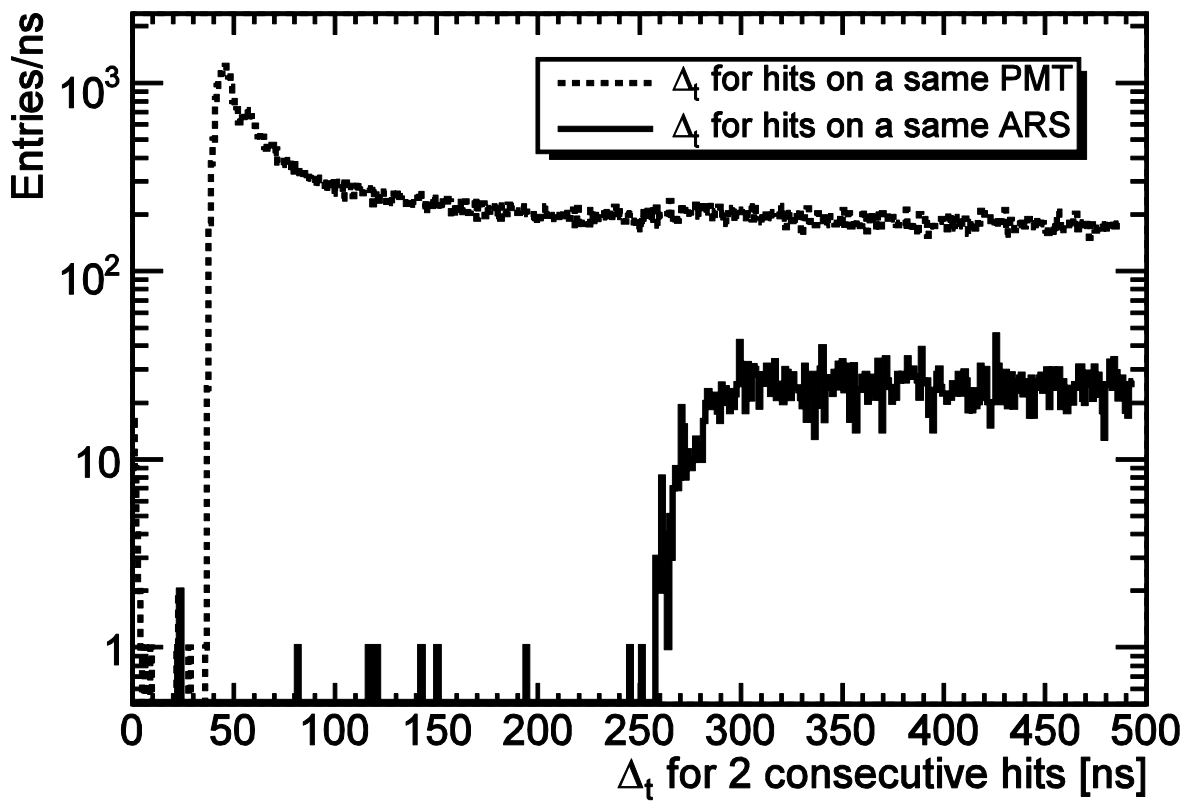

Figure 21. Time difference between two consecutive hits. Upper curve: two hits in the same PMT. Lower curve: two hits in the same ARS.

The chip dead time is the convolution of three independent contributions:

(1) The dead time associated with the TVC and charge measurement. When an ARS is processing an event, it cannot process a new one. This dead time depends on the integration gate and on the writing time into the pipeline. It can take values between 120 ns and 500 ns depending on the ARS configuration. In the setup used in the experiment it is around 250 ns. The token ring protocol chaining the two ARSs chips minimizes this dead time, so the minimum time difference between two consecutive hits in the same PMT is 38 ns, due to the integration gate and the token ring delay.

(2) The dead time associated with the sampler, relevant in PSD or WF mode. When an event is declared “WF”, the ARS stores all the samples until they are read or erased. It takes 145 μs to transmit the data of the anode WF and of the sampled clock. The rate of WF events from accidental coincidences of single pulses is expected to be of some hundreds of Hz.

(3) The last dead time concerns the saturation of the pipeline memory. It depends on the readout clock frequency used to unload cell information and on the event rate. With a readout frequency of 25 MHz and a 48-bit SPE event, the 16 cells of the pipeline allow a writing rate without loss of SPE events up to 490 kHz mean rate with Poisson distribution. However, in order to ensure good transmission conditions of the data from the DAQ card data taking is suspended when the background rate exceeds 400 kHz.

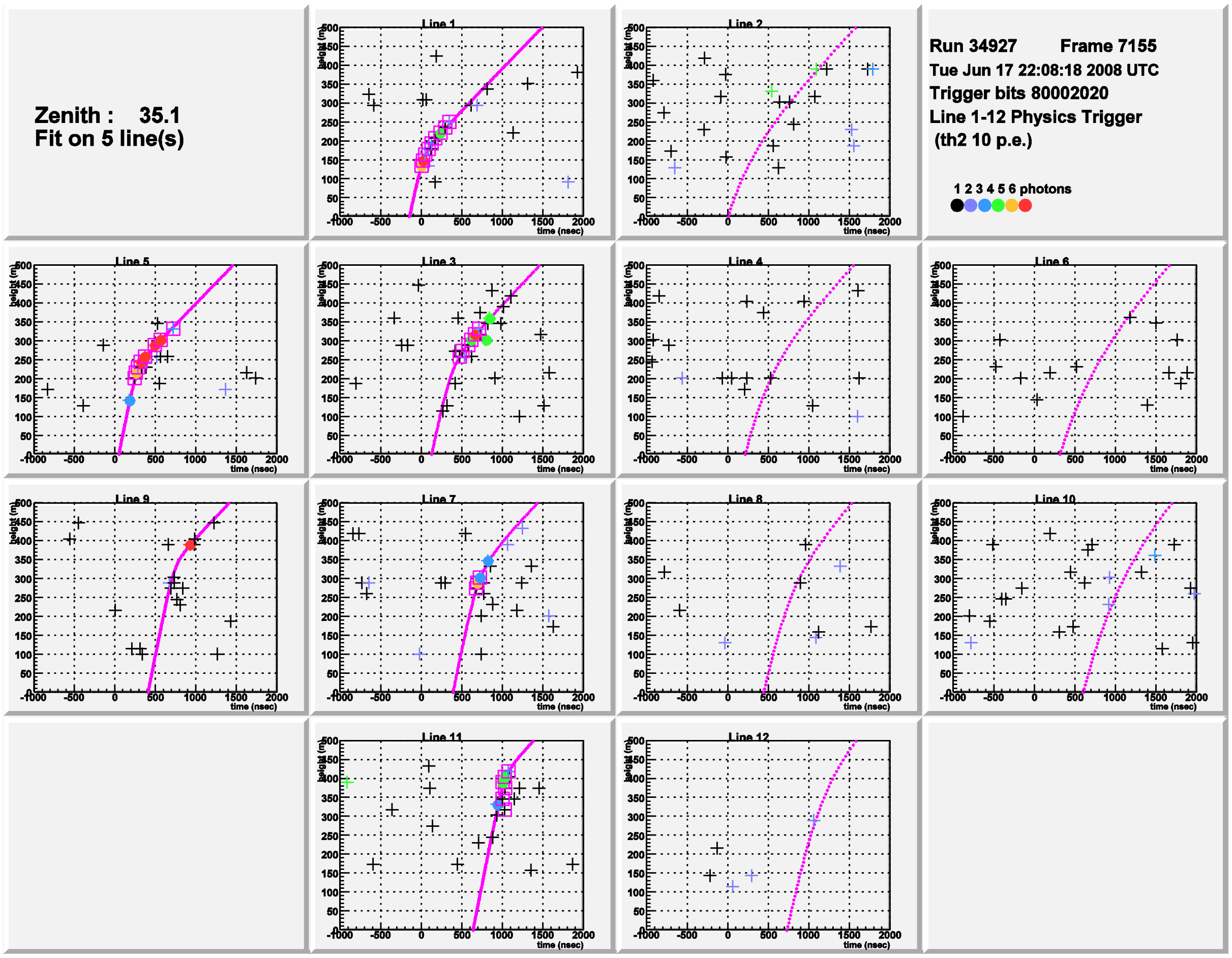

Figure 22 Time-altitude (t-z) plots of a neutrino-induced muon reconstructed as up-going at 34.8° from the vertical with the online reconstruction algorithm. Each box corresponds to the time-altitude plot of the hits observed on a line. The position of the 12 boxes approximately reflects the relative position of the lines on the seabed. The altitude and the time of the hits from the Cherenkov light cone are related by a hyperbola. Hits that have been used in the fit are drawn as squares. The colour code indicates the charge measurement as shown in the caption. Higher intensity of the track line indicates closer distance of the reconstructed track to the detector line.

Figure 20 shows the timeline of two consecutive events and the corresponding occupancy in the front end for the cases of SPE only and mixed (PSD) operation. The dead time between two consecutive events can be checked using various methods. Using the optical background the dead time is easily measured as shown in Figure 21 for the simple SPE mode. The minimum time differences between two consecutive hits, 38 ns in the same PMT and 250 ns in the same ARS, are observed. Another measurement of the dead time, using reconstructed down-going cosmic muons, is presented in [19].

The dead time of the electronics does not introduce any significant inefficiency in track reconstruction:

- In the case of two consecutive photons close in time: the main effect of the 25 ns integration gate occurs when a signal photon falls within the gate opened by another photon. If the first photon is from the track the two hits are merged in the reconstruction; if it is from background the timing of the signal photon will be wrong, so it will be lost or will enter in the fit with a wrong timing. When a signal photon falls within the 13 ns needed for the token ring transmission it is lost. The probability of a random hit coinciding with a signal hit within t= 25+13= 38 ns ranges from $2.3 \times 10^{-3}$ at 60 kHz background to $1.6 \times 10^{-3}$ at the highest acceptable background, 400 kHz.
- The probability of hit loss in the case of three consecutive photons, with both ARSs busy processing, is lower.

In both cases the weak proportion of missing or wrong hits does not affect the fit results. At high rates the track reconstruction algorithms limit the performance rather than the front end.

# 7 Conclusions and outlook

The front-end electronics and in particular the ARS fulfilled the specifications for the physics purposes of the ANTARES detector. An advantage of this chip, given all its functionalities, is the low power consumption, 190 mW at 5 V power supply. Thanks to the monorail power supply it is easy to integrate the chip in a system. The large number of integrated functions and the asynchronous electronics operation, made this project very challenging. The ARS chip was produced with a good yield and has proved reliable in data taking. The required timing resolution better than 500 ps and the fast sampling of signal waveforms have been achieved.

The success of the ANTARES electronics system is demonstrated by the high quality of the event reconstruction, an example of which is shown in Figure 22.

# Acknowledgements

The authors acknowledge the financial support of the funding agencies: Centre National de la Recherche Scientifique (CNRS), Commissariat à l'énergie atomique et aux énergies alternatives (CEA), Commission Européenne (FEDER fund and Marie Curie Program), Région Alsace (contrat CPER), Région Provence-Alpes-Côte d'Azur, Département du Var and Ville de La Seyne-sur-Mer, France; Bundesministerium für Bildung und Forschung (BMBF), Germany; Istituto Nazionale di Fisica Nucleare (INFN), Italy; Stichting voor Fundamenteel Onderzoek der Materie (FOM), Nederlandse organisatie voor Wetenschappelijk Onderzoek (NWO), the Netherlands; Council of the President of the Russian Federation for young scientists and leading scientific schools supporting grants, Russia; National Authority for Scientific Research (ANCS), Romania; Ministerio de Ciencia e Innovación (MICINN), Prometeo of Generalitat Valenciana (GVA) and MultiDark, Spain. We also acknowledge the technical support of Ifremer, AIM and Foselev Marine for the sea operation and the CC-IN2P3 for the computing facilities and the financial support of the French Agence National de la Recherche under the contract number ANR-08-JCJC-0061-01.